\documentclass[aps,nofootinbib,twocolumn,superscriptaddress,floatfix,showpacs]{revtex4-1}
\usepackage[USenglish]{babel}
\usepackage{epsfig}
\usepackage{amssymb}
\usepackage{amsmath}
\usepackage{bm}
\usepackage{hyperref}
\usepackage{enumerate}

\begin{document}

\title{Functional renormalization group for commensurate antiferromagnets:\\ Beyond the mean-field picture}
\author{Stefan A. Maier}
 \email{smaier@physik.rwth-aachen.de}
\affiliation{Institute for Theoretical Solid State Physics, RWTH Aachen University, D-52056 Aachen, Germany}
\author{Andreas Eberlein}
\affiliation{Max Planck Institute for Solid State Research, D-70569 Stuttgart, Germany}
 \author{Carsten Honerkamp}
\altaffiliation{member of JARA -- FIT, J\"{u}lich-Aachen Research Alliance -- Fundamentals of Future Information Technology}
\affiliation{Institute for Theoretical Solid State Physics, RWTH Aachen University, D-52056 Aachen, Germany}

\date{\today}

\begin{abstract}
We present a functional renormalization group (fRG) formalism for interacting fermions on lattices that captures the flow into states with commensurate spin-density-wave order. During the flow, the growth of the order parameter is fed back into the flow of the interactions and all modes can be integrated out. This extends previous fRG flows in the symmetric phase that run into a divergence at a nonzero RG scale, i.e., that have to be stopped at the ordering scale. We use the corresponding Ward identity to check the accuracy of the results. We apply our method to a model with two Fermi pockets that have perfect particle-hole nesting. The results obtained from the fRG are compared with those in random-phase approximation.
\end{abstract}

\pacs{05.10.Cc,75.10.Lp,75.50.Ee}

\maketitle

 \section{Introduction}
States with broken symmetry play an important role in our understanding of interacting electron systems. Among many possibilities, electron spin magnetism is certainly one of the most important examples. \cite{auerbach,schollwoeck} For strongly interacting, localized electrons, pure spin models can be used to study magnetic order and the corresponding excitations. If the interactions are weaker and the electrons near the Fermi level are delocalized, magnetism can also occur as an
 instability that modifies the single-particle excitation spectrum. Such a behavior is even more likely in layered, quasi-two-dimensional systems,
 where the electronic bands can cause sharp peaks in the density of states and nesting of sizable regions in the Brillouin zone.
 In these cases, characteristic length scales are typically larger and energy scales lower than in the localized moment systems,
 and details of the band structures play a role. This sets strong limitations to the applicability of theoretical methods.
  For the systems in question, theoretical insight can be gained from methods in the realm of random-phase approximation (RPA) or mean-field theory (MFT).
 \emph{Ab initio} techniques such as the spin-resolved density functional theory \cite{spin-DFT} fall into the same class, regarding the treatment of fluctuations 
 and competing effects. Usually, information on the type and strength of the ordering can be obtained from these methods,
 while, in particular, the latter is often overestimated.
 For drawing a more precise and rounded picture, the functional renormalization group (fRG) is expected to become a useful tool. 
 So far, in the context of interacting electron systems, RG methods have been employed a lot to determine the leading ordering tendencies (for reviews, see Refs.~\onlinecite{metzner-rmp,multi-orbital-review}). The strength of these methods lies in their unbiased treatment of fluctuations.
 More precisely, they sum up all important fluctuations in different channels together, instead of singling 
 a particular one out as the MFT. 
 However, the information from the RG is usually restricted to the type of order and a rough estimate of the corresponding energy scale.
 At this energy scale, the RG flow has to be stopped as a runaway-flow of the coupling constants spoils the applicability of the flow equations. 
 A real improvement would be to avoid the unphysical (but meaningful) runaway-flow and continue the RG flows into the low-energy regime of broken symmetry. Then, all scales could be integrated out and a  renormalized band structure together with a fluctuation-corrected value of the ordered moment would be obtained.

The development of fRG methods that can flow towards a symmetry-broken state has already seen some initial stages. In a number of papers, simplified
 mean-field models with infinitely long-ranged interactions have been studied. \cite{brokensy_SHML,gersch_cdw,gersch_firstorder,me-AFM-MF} 
 For these models, MFT holds exactly due to the specific form of their interaction terms, and this exact solution can be recovered within a fRG framework.
More generic interactions were treated by R.~Gersch \emph{et al.\ }in Ref.~\onlinecite{gersch_superfluid} for the attractive Hubbard model, and in a more recent study, by one of us and W.~Metzner for the repulsive Hubbard model. \cite{eberlein-repu} 

In both cases, the order parameter that developed during the flow was of superconducting type. In particular, the aforementioned work on the
 repulsive Hubbard model continues a line of RG studies (for a review, see Ref.~\onlinecite{metzner-rmp}) showing a leading pairing instability in the $d$-wave channel for the Hubbard model and gives precise values for the gap magnitude as a function of the model parameters. Also more recently, two of us have worked
 out the analogous flows for the case of spin-density-wave (SDW) magnetic ordering in two dimensions, \cite{me-AFM-MF} again first for a model that is exactly solvable in MFT. In this work, we give the extension to the more general case of short-range initial interactions, choosing a model with two pockets that are perfectly nested.

As the main outcome of our work,
 we find that
 the extension into the SDW regime gives useful results for our test model and can hence be applied to more sophisticated cases.
 The scheme allows for determining corrections to the mean-field picture, e.g.\ for the size and parameter dependence of the magnetic order parameter. In addition to obtaining numerical values for these corrections, one can also gain insight about the impact of other collective channels on them. In our case, the charge-density wave (CDW)
 and the singlet-pairing  channel have a substantial impact on the corrections to the SDW order parameter.

The paper is organized as follows. In Sec.~\ref{sec:AFM-ch-dec}, we present a channel-decomposed one-particle irreducible (1PI) fRG scheme for antiferromagnets.
 After reviewing the fRG flow for charge-conserving theories, we give a parametrization of the interaction that exploits the residual
 spin symmetry for collinear magnetic ordering in Sec.~\ref{sec:param-gen}.
 For the special case of antiferromagnetic ordering, we then discuss the interaction terms allowed by the remaining symmetries.
 In Sec.~\ref{sec:AFM-fleq}, we give an approximate parametrization of the 1PI functional, which plays the role of an effective action.
 In the course of these approximations, we resort to an exchange parametrization within a channel decomposition, and we neglect the breaking of the discrete
 time-reversal and translational symmetries. At the one-particle level, however, the breaking of these symmetries will be retained. The underlying formalism for the channel
 decomposition and the exchange parametrization is laid out in Appendices~\ref{sec:ch-dec} and \ref{sec:AFM-ex-para}, respectively.
 We then give the corresponding fRG flow equations, which are derived in Appendix~\ref{sec:deriv-ex-para}.
 In Sec.~\ref{sec:rpa}, we discuss these flow equations in RPA and observe that the mean-field gap equation is recovered exactly
 from the fRG.

 In Sec.~\ref{sec:appl}, we consider the fRG flow into the SDW phase of a simple two-pocket model \cite{chubukov} at perfect nesting.
 After discussing this model, we give details of our implementation in Sec.~\ref{sec:chubu-impl}.
 The numerical solution of the fRG flow equations is discussed in Sec.~\ref{sec:numflow}. We then comment on the fulfillment of the global SU(2) Ward identity
 for the gap in Sec.~\ref{sec:WI-fulfill}. Finally, we compare the gaps obtained from fRG to MFT  in Sec.~\ref{sec:comp-MF} and
 show that, despite the relatively small renormalizations of the mean-field results, the coupling of different interaction channels plays an important role 
 in the flow equations. Therefore, the (subleading) charge-density wave and pairing channels need to be included in order to obtain reasonable results from the
 fRG.
\section{Method} \label{sec:AFM-ch-dec}
 In this section, we present a 1PI fRG scheme for commensurate antiferromagnets. While it will be applied to a simple two-pocket model in Sec.~\ref{sec:appl},
 it is not solely designed for that particular model. It may therefore be carried over to other models of interest such as single- and multiband Hubbard models.
\subsection{General context} \label{sec:gen-context}
In this paper, we will study models with actions of the form 
 \begin{align} \notag
 \mathcal{A}[\bar{\psi},\psi] &= \sum_\sigma \int \! dk \, dk' \, \bar{\psi}_\sigma (k) \, C^0_\sigma (k,k') \, \psi_\sigma (k')\\  \label{eqn:action}
 & \quad + \mathcal{A}^{(4)} [\bar{\psi},\psi]
 \end{align}
and interactions
\begin{align} \notag
 \mathcal{A}^{(4)} [\bar{\psi},\psi] & = \frac{1}{4} \int \! d \xi_1 \dots d \xi_4 \, f(\xi_1,\xi_2,\xi_3,\xi_4) \\ \label{eqn:form-int}
 & \qquad \times \bar{\psi} (\xi_1) \, \bar{\psi} (\xi_2) \, \psi (\xi_3) \, \psi (\xi_4) \, ,
\end{align}
with Grassmann fields $ \psi $ and variables $\xi_i$ that include the spin-projection $\sigma_i$, momentum $\mathbf{k}_i$ and Matsubara frequency $k_{0,i}$. (In Sec.~\ref{sec:chubu}, we will study a two-dimensional model, but the formalism presented here is applicable for arbitrary lattice dimensions.)
 The two latter quantum numbers are also collected in a generalized momentum $k = (k_0, \mathbf{k})$.
$C^0_\sigma (k,k')$ is the inverse free propagator of the theory, and $ f(\xi_1,\xi_2,\xi_3,\xi_4)$ accounts for the interaction.
 The functional renormalization group (fRG) flow then describes the change of the one-particle irreducible (1PI) vertices when the modes summed over in the quadratic part in Eq.~(\ref{eqn:action}) are integrated out in a continuous way. In this context, we call $\mathcal{A}$ in Eq.~(\ref{eqn:action}) the initial action of the system. 
%For a recent review on the fRG, see Ref.~\onlinecite{metzner-rmp}. 

The terms that are allowed, i.e. nonzero, in Eq.~(\ref{eqn:action}) can be analyzed by considering the symmetries of the system. 
This usually simplifies the study a lot. The translational symmetry on the lattice and in imaginary time renders the quadratic part diagonal in ${k}$, and makes the interaction term only depend on three instead of four $k$s. Furthermore, spin-rotational symmetry allows for replacing the spin-index-afflicted function $f(\xi_1,\xi_2,\xi_3,\xi_4) $  by a coupling function $V(k_1,k_2,k_3)$ as in Ref.~\onlinecite{salm_hon_2001}.
However, for a system with spontaneously broken symmetries such as an antiferromagnet, new terms that are absent in the initial action may occur at lower energy scales, or, in other words, during the RG flow.  This issue will be covered in the next subsection.

In a fRG flow, a dependence on a cutoff scale $\lambda$ is introduced in the quadratic part of the action. 
The regularization scheme underlying this replacement
$ C_\sigma (k,k') \to C^\lambda_\sigma (k,k') $
does not need to be specified \emph{a priori}.
As laid out in Ref.~\onlinecite{salm_hon_2001},
the flow of the 1PI self-energy is governed by the equation
\begin{equation*}
 \partial_\lambda \Sigma (\xi_1,\xi_2) = \int \! d \eta_1 \, d \eta_2 \, S(\eta_2,\eta_1) \,
  f(\xi_1,\eta_1,\eta_2,\xi_2) \, ,
\end{equation*}
where $ S $ denotes the single-scale propagator
\begin{align} \notag
  S  (\xi_1,\xi_2) & = \partial_\lambda G (\xi_1,\xi_2) - \int \! d\eta_1 \, d\eta_2 \, G (\xi_1,\eta_1) \\  \label{eqn:def-sscale}
 & \quad \times \left[ \partial_\lambda \Sigma (\eta_1,\eta_2) \right] G (\eta_2,\xi_2) \, ,
\end{align}
and $ G (\xi_1,\xi_2) $ the \emph{full} propagator.
In the Katanin truncation \cite{katanin_trunc} of the 1PI fRG scheme, 
the flow of the interaction for a charge-conserving theory is given by
\begin{align*}
 \partial_\lambda  f (\xi_1,\xi_2,\xi_3,\xi_4) &= F_{\rm pp} (\xi_1,\xi_2,\xi_3,\xi_4) \\ & \quad +
F_{\rm ph} (\xi_1,\xi_2,\xi_3,\xi_4) \\ & \quad -
F_{\rm ph} (\xi_1,\xi_2,\xi_4,\xi_3) \, ,
\end{align*}
where the right-hand side consists of particle-particle and particle-hole diagrams,
\begin{align*}
 F_{\rm pp} (\xi_1,\xi_2,\xi_3,\xi_4) & = \frac{1}{2} \int \!\! d \eta_1 \, d \eta_2 \, d \eta_3 \, d \eta_4 
\,  f(\xi_1,\xi_2,\eta_2,\eta_3) \,  \\ & \quad \times   \left[ \partial_\lambda G(\eta_2,\eta_1) \, G(\eta_3,\eta_4) \right] \\
 & \quad \times f (\eta_4,\eta_1,\xi_3,\xi_4)
\end{align*}
and
\begin{align*}
 F_{\rm ph} (\xi_1,\xi_2,\xi_3,\xi_4) & = - \int \!\! d \eta_1 \, d \eta_2 \, d \eta_3 \, d \eta_4 \,
 f (\eta_4,\xi_2,\xi_3,\eta_1)  \\ & \quad \times 
 \left[ \partial_\lambda G(\eta_1,\eta_2) \, G(\eta_3,\eta_4) \right]\\ & \quad \times 
 f (\xi_1,\eta_2,\eta_3,\xi_4) \, ,
\end{align*}
respectively.
While three-particle and higher interaction terms are not taken into account explicitly, contributions from the three-particle vertex
are partly included in $ F_\mathrm{pp}$ and $F_\mathrm{ph}$.
\subsection{Parametrization for collinear spin ordering} \label{sec:param-gen}
 In an antiferromagnet (AF), at least two different symmetries are broken spontaneously. For one thing, translational invariance is reduced in an AF phase as the magnetic unit cell defined by the ordering pattern is larger than the unit cell given by the lattice structure. For \emph{commensurate} ordering, the magnetic cell volume is an integer multiple of the lattice unit cell. In this work, the unit cell of the symmetry-broken state will be twice as large as in the symmetric phase.
 Moreover, a collinear AF state breaks the SU(2) symmetry by spontaneously selecting a preferred axis for the alignment of the spins. Let this be the $z$ axis.
 Then, there are still remnants of the SU(2) symmetry. Namely, the system stays invariant under spin rotations in the $xy$ plane.
 
 We work with Green's functions and one-particle irreducible (1PI) vertices as basic elements.
 These objects contain a wealth of information and need to be parametrized in an efficient way.   
 Such a parametrization should therefore take the remaining symmetry for collinear spin ordering into account. In Ref.~\onlinecite{me-AFM-MF},
 we have already given such a parametrization for charge-conserving theories, which we now briefly recapitulate.
 (In a real-space description, a similar parametrization was used in Ref.~\onlinecite{xxz_spinRG}.)
The elements of the remaining
spin symmetry group $ \mathrm{U}_z (1) $ are $ U(\varphi)= e^{i \varphi \tau^z} $ with arbitrary real $ \varphi $ and the third Pauli matrix
 $\tau^z$ acting on the spin space spanned by spin up and spin down with respect to the $z$ axis. 
 The action of a system with collinear spin symmetry stays invariant if the Grassmann fields are transformed by an element of $ \mathrm{U}_z (1) $.

 Then, for the magnetic ordering along the $z$ axis, the one-particle Green's function only has diagonal entries in spin space. It can hence
 be split into a spin-reversal symmetric and a spin-reversal antisymmetric part according to
\begin{equation*}
  G_{\sigma_1,\sigma_2} = G_1 \delta_{\sigma_1,\sigma_2} + G_z \tau^z_{\sigma_1,\sigma_2} = G_{\sigma_1} \, , 
\end{equation*}
with the spin indices  $\sigma_i$ taking on the values $\uparrow$ or $\downarrow$. 
Consider next a two-particle interaction $\mathcal{A}^{(4)}$ of the form given in Eq.~(\ref{eqn:form-int}).
 The $\mathrm{U}_z (1)$  symmetry now restricts the spin-dependence of the $f$ to the form
\begin{align*}
f (\xi_1,\xi_2,\xi_3,\xi_4) & =V_\uparrow (k_1,k_2,k_3,k_4) \, \delta_{{\boldsymbol \sigma},\uparrow \uparrow \uparrow \uparrow}  \\
 & \quad  + V_\downarrow (k_1,k_2,k_3,k_4) \, \delta_{{\boldsymbol \sigma},\downarrow \downarrow \downarrow \downarrow} \\ 
& \quad + V_{\uparrow\downarrow} (k_1,k_2,k_3,k_4) \, \delta_{{\boldsymbol \sigma},\uparrow \downarrow \uparrow \downarrow} \\
 & \quad  - V_{\uparrow\downarrow} (k_1,k_2,k_4,k_3) \, \delta_{{\boldsymbol \sigma},\uparrow \downarrow \downarrow \uparrow}\\
  & \quad + V_{\uparrow\downarrow} (k_2,k_1,k_4,k_3) \, \delta_{{\boldsymbol \sigma},\downarrow \uparrow \downarrow \uparrow} \\
 & \quad  - V_{\uparrow\downarrow} (k_2,k_1,k_3,k_4) \, \delta_{{\boldsymbol \sigma},\downarrow \uparrow \uparrow \downarrow} \, .
\end{align*}
 Due to the antisymmetry property of $f (\xi_1,\xi_2,\xi_3,\xi_4) $, $ V_\uparrow $ and $ V_\downarrow $ are antisymmetric under $ k_1 \leftrightarrow k_2 $
and $ k_3 \leftrightarrow k_4 $, whereas the Pauli principle  does not impose a constraint on $ V_{\uparrow\downarrow}$. 

A global $ \mathrm{SU} (2) $ Ward identity for the self-energy
 can be derived in analogy to the U(1) case in Ref.~\onlinecite{brokensy_SHML}, Eq.~(85). One obtains
\begin{align}  \notag
 C_z (k_1,k_2) -  C^0_z (k_1,k_2) & = 
 - \int \!\! d p_1 \,d p_2 \,d p_3 \,d p_4 \, C^0_z (p_1,p_2) \\ \notag & \qquad \times
 G_\downarrow (p_2,p_3) \, G_\uparrow (p_4,p_1) \\ \label{eqn:genWI} 
 & \qquad \times V_{\uparrow \downarrow} (k_1,p_3,p_4,k_2) \, ,
\end{align} 
where $ C_z $ and $ C^0_z $ denote the spin-antisymmetric part of the inverse of the full and the bare one-particle propagator, respectively. Note that $ G_\sigma $  represents the full propagator and that $ V_{\uparrow \downarrow} $ enters as a renormalized interaction. We will make use of the Ward identity later on, as a second way to assess the self-energy in addition to obtaining it from the renormalization group flow.

\subsection{Two-particle interaction terms in an antiferromagnet}
In order to get some intuition for the particularities of the channel-decomposed flow equations in the presence of antiferromagnetic
 ordering, let us first discuss
 processes mediated by some kind of exchange boson that comply with the remaining symmetries.
 In addition to the contributions that are already present in the symmetric phase, there will be processes that violate the translational
 or $ \mathrm{SU} (2) $ symmetries or both.

 Let us start with discussing the Nambu-index dependence of the interaction.
In the case of commensurate AF, the renormalized interaction is only invariant under translations by an even number of sites. In momentum space, the ordering vector $\mathbf{Q}$ then corresponds to half a reciprocal lattice vector. 
Accordingly, the coupling functions can be decomposed into a momentum-conserving  part $ V_{\dots}^\mathrm{c} $ and a nonconserving part $ V_{\dots}^\mathrm{nc} $, which is generated during the flow. We then have 
\begin{align*}
 V_{\dots}  (k_1,k_2,k_3,k_4) & = V_{\dots}^\mathrm{c} (k_1,k_2,k_3) \, \delta \left( k_1 + k_2 - k_3 - k_4 \right) \\
 & \quad +V_{\dots}^\mathrm{nc} (k_1,k_2,k_3) \\ & \qquad \times \delta \left( k_1 + k_2 - k_3 - k_4  +Q \right) \, ,
\end{align*}
with $Q = (0,\mathbf{Q}) $.
In Nambu representation with pseudo-spinors
\begin{equation} \label{eqn:Nambu-spinor}
\bm{\Psi}_\sigma (k) = \left( \begin{array}{l} \psi_\sigma (k) \\ \psi_\sigma (k+Q) \end{array} \right) \, ,
\end{equation}
the interaction can be parametrized in the same way as in the conventional representation with coupling functions $ V_{\dots} (k_1,k_2,k_3,k_4)$ using coupling functions $ W_{\dots} (K_1,K_2,K_3,K_4) $, where $ K_i = (k_i,s_i) $ with Nambu indices $ s_i$.

For an even number of equal Nambu indices $ s = \pm 1 $, the interaction $ W_{\dots}$ in Nambu representation then corresponds to $ V_{\dots}^\mathrm{c}$ and  to $ V_{\dots}^\mathrm{nc}$ for an odd number of equal Nambu indices. Thus, one has
\begin{equation*}
 W_{\dots} (K_1,K_2,K_3,K_4) = \tilde{\delta}_{\left\{ k_i \right\}} \, V_{\dots}^\mathrm{c} (\varkappa_1,\varkappa_2,\varkappa_3) 
\end{equation*}
 for even $\sum_i \frac{s_i}{2} $, and
\begin{equation*}
 W_{\dots} (K_1,K_2,K_3,K_4) = \tilde{\delta}_{\left\{ k_i \right\}} \, V_{\dots}^\mathrm{nc} (\varkappa_1,\varkappa_2,\varkappa_3) 
\end{equation*}
otherwise, where the physical momenta are denoted by $ \varkappa_i = k_i + \left( 1 - s_i \right) Q /2 $.
In these formulae, the momenta $ \mathrm{k}_i $ are restricted to half the BZ, and therefore $ \tilde{\delta}_{\left\{ k_i \right\} } $ ensures momentum conservation only up to multiples of the ordering vector $ \mathbf{Q} $.

 In contrast to the discrete translational invariance, the $ \mathrm{SU} (2) $ symmetry is a continuous one. Its spontaneous breaking is therefore accompanied
 by the emergence of 
 massless Goldstone modes. In a purely fermionic language, this will be reflected by the divergence of some contributions to
 the two-particle interaction in the limit of a vanishing seed field, as for a broken U(1) 
 symmetry. \cite{brokensy_SHML,gersch_superfluid,eberlein_param,eberlein-sssu}
 The radial mode, however, will have a mass and therefore the corresponding contributions to the interaction remain regular for a vanishing
 seed field. Let us now define the fermionic spin-density-wave (SDW) bilinear
\begin{equation*}
 S_i^{s_1 s_2} (l) = \sum_{\sigma \sigma'} \int \! d'k \, \bar{\Psi}^{s_1}_\sigma (k+l/2) \, \tau^i_{\sigma \sigma'} \Psi^{s_2}_{\sigma'} (k-l/2)  \, ,
\end{equation*}
 where $ \tau^i $ with $i=x,y,z$ denotes a Pauli matrix and there the momentum integration only runs over the magnetic BZ, which is
 indicated by the prime in the measure $d'k$.
 Note that the Nambu indices are treated as some kind of flavor quantum numbers here.
 For $l=0$,
the physical SDW transfer momentum amounts to $ \mathbf{l} $ if $ s_1 = s_2 $ and to $ \mathbf{l} + \mathbf{Q} $ if $ s_1 \neq s_2 $.
 In the spirit of a gradient expansion around the center of the magnetic BZ, this picture still holds in an approximate sense also for $ l \neq 0$.

\begin{figure}
\centering
\includegraphics[width=.8\linewidth]{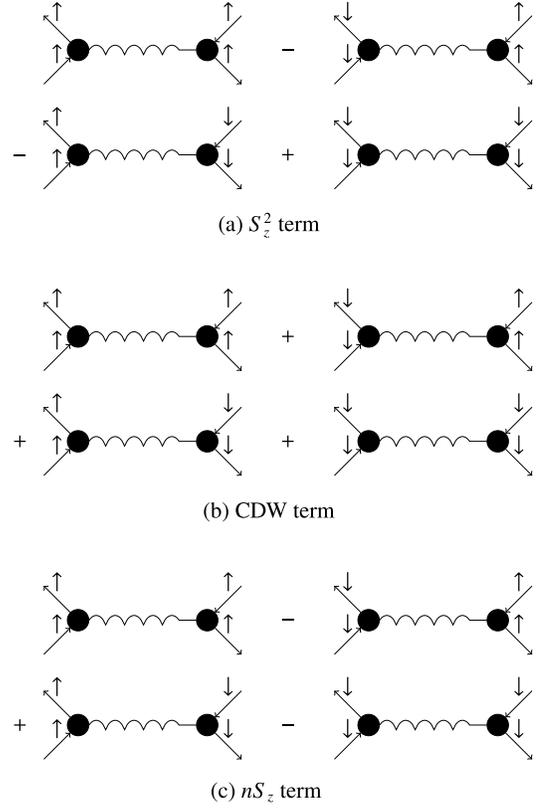}
\caption{CDW, $S_z^2$, and spin-anomalous $S_z n$ terms in a boson-exchange picture.} \label{fig:new-props}
\end{figure}

In a boson-exchange picture, the Goldstone and radial vertices then correspond to $ S_x^2 + S_y^2 $ and $ S_z^2 $ terms,
 respectively. 
 [For a pictorial representation of the $S_z^2$ term, see Fig.~\ref{fig:new-props}(a).]
If the $ \mathrm{SU} (2) $ symmetry is broken, those terms differ.
 Even though, they both are still invariant under a reversal of the spin-projection quantum number. 
In the following, we will call such contributions to the interaction spin-normal.
 The remaining $ \mathrm{U}_z (1) $ symmetry also allows for spin-anomalous terms of the form $ S_x S_y $.
 So in a channel decomposition of the fRG flow equations, the magnetic channel of Refs.~\onlinecite{husemann_09,HGS-freq-dep,giering-new}
 should split into radial and Goldstone as well as spin-anomalous contributions.
 In addition, there will be charge-density wave (CDW) contributions of the form $n^2$ (see Fig.~\ref{fig:new-props}(b)), where
\begin{equation*} 
 n^{s_1 s_2} (l) = \sum_{\sigma } \int \! d'k \, \bar{\Psi}^{s_1}_\sigma (k+l/2) \, \Psi^{s_2}_{\sigma} (k-l/2)
\end{equation*} 
 denotes the CDW bilinear. Also spin-anomalous $S_z n$ contributions as depicted in Fig.~\ref{fig:new-props}(c) are allowed.

 So far, we have discussed different particle-hole exchange processes. One may now wonder  whether the breaking of the
 $ \mathrm{SU} (2) $ symmetry has also nontrivial consequences for contributions to the interaction induced by the exchange of virtual Cooper pairs.
 One may intuitively expect that there will be singlet- and triplet-pairing terms, but
 this classification does not apply in a strict sense if the $ \mathrm{SU} (2) $ symmetry is broken.
 This can be seen by considering the Cooper-pair bilinear
\begin{equation*}
 \phi_{\sigma_1 \sigma_2}^{s_1 s_2} (l) = \int \! d'q \,  \bar{\Psi}_{\sigma_1}^{s_1} \left( l/2 + q \right)
\bar{\Psi}_{\sigma_2}^{s_2} \left( l/2 - q \right) \, ,
\end{equation*}
 which equals $-\phi_{\sigma_2 \sigma_1}^{s_2 s_1} (l) $ due to the Pauli principle. 

In the presence of $ \mathrm{SU}(2) $ spin rotation
invariance, Cooper pairs can be classified as singlet and triplet pairs,
which do not get mixed under spin rotations.
 The singlet-pair bilinear, for example, reads as
\begin{equation*}
 \phi_\mathrm{sing}^{s_1 s_2} (l) =  \frac{i}{\sqrt{2}} \sum_{\sigma \sigma'} \tau^y_{\sigma \sigma'} \phi_{\sigma \sigma'}^{s_1 s_2} (l) \, .
\end{equation*}
 If the $\mathrm{SU} (2)$ rotation
invariance is broken as in the case of antiferromagnetic order, the triplet
component splits and $ \phi_{\uparrow \uparrow} $, $ \phi_{\downarrow \downarrow} $ and $ \left( \phi_{\uparrow \downarrow} + \phi_{\downarrow \uparrow} \right) / \sqrt{2} $ are not degenerate. Thus, in this case, $\phi$ can be decomposed in
four $ \mathrm{U}_z(1) $ invariant parts.
This observation has also been made by D.~Scherer \emph{et al.} in the context of the Kitaev-Heisenberg model. \cite{fRG-Kitaev}
 In the following,
 the terms `singlet' and `triplet pairing' will only be used for $ \mathrm{SU} (2) $ invariant contributions to the interaction. Conversely, $ \mathrm{SU} (2) $-breaking  contributions in the
 pairing channels will be called `anomalous pairing terms'.
Having qualitatively discussed the consequences of the broken $\mathrm{SU}(2)$ and
translational symmetries, we are now in a position where we can
formulate our approximate parametrization of the effective action in an
antiferromagnet. The parametrization is based on a decomposition of the
vertex in interaction channels, which is presented in Appendix \ref{sec:ch-dec}.

\subsection{Approximate parametrization of the effective action and fRG flow equations} \label{sec:AFM-fleq}

 In order to reduce the computational cost of our fRG flow, we now resort to an approximate parametrization of the effective action
\begin{equation*}
\Gamma [\bar{\Psi},\Psi] = \sum_\sigma \int \! d'k \, \bar{\Psi}_\sigma (k) \, \mathbf{C}_{\sigma} (k) \, \Psi_\sigma (k)
 +\Gamma^{(4)} [\bar{\Psi},\Psi] \, ,
\end{equation*}
 in which only the most important renormalizations are retained.
 It contains the inverse $\mathbf{C}$ of the one-particle propagator in the quadratic part, and, in agreement with the flow equations in Sec.~\ref{sec:gen-context},
 the interactions are truncated after the two-particle term $\Gamma^{(4)}$.
 At the two-particle level, anomalous contributions breaking discrete symmetries seem to be of minor importance in an antiferromagnet. At the one-particle level, however, such anomalous terms are retained, and we allow for the breaking of the continuous SU(2) symmetry in our parametrization of the interaction. 
 While the former is essential for obtaining a nonzero AF order parameter, the latter reflects the physics of the Goldstone theorem.

 Before writing an ansatz for the effective action and giving the corresponding fRG flow equations, let us first elaborate on the role
 of the breaking of discrete symmetries in an antiferromagnet.
 While a nonzero staggered magnetization obviously implies the breaking of translational symmetry, 
 both a time-reversal operation and a translation by one primitive lattice vector only change the sign of the order parameter. This sign, however, is fixed
 by some arbitrary convention and does \emph{not} reflect any macroscopic property of the system. Of course, we have to fix the sign of the AF gap
 at the one-particle level, but it appears likely that the breaking of discrete symmetries flipping this sign can be neglected at the two-particle level.%
\footnote{Note that, for a time-reversal operation, the situation is different in a \emph{ferromagnet}. Namely, the sign of the order parameter does have
some macroscopic content in that case and, consequently, time-anomalous interaction terms should be kept.}
 In the following, we will do so and the Ward identity (WI) for the gap will serve as a measure of the quality of our approximations.

 In addition to dropping time-reversal breaking interaction terms, we now resort to an exchange parametrization
  of the interaction within a channel decomposition. This decomposition is discussed in Appendix~\ref{sec:ch-dec} in detail, and the exchange parametrization
 in Appendix~\ref{sec:AFM-ex-para}. The interaction then takes on the form
\begin{equation*}
 \begin{aligned}
 \Gamma^{(4)} [\bar{\Psi},\Psi] &= \mathcal{A}^{(4)} [\bar{\Psi},\Psi]\\
 & \quad + 2 \sum_{\{s\}} \int \! d'l \,  \phi_\mathrm{sing}^{s_1 s_2} (l) \, {\phi_\mathrm{sing}^{s_3 s_4}}^\ast (l) \, D^{\{s\}} (l)\\
 & \quad + \sum_{\{s\}} \int \! d'l \,  n^{s_1 s_4} (l) \, n^{s_2 s_3} (l) \, N^{\{s\}} (l)\\
 & \quad + \sum_{\{s\}} \int \! d'l \,  S_x^{s_1 s_4} (l) \, S_x^{s_2 s_3} (l) \, M_{xy}^{\{s\}} (l)\\
 & \quad + \sum_{\{s\}} \int \! d'l \,  S_y^{s_1 s_4} (l) \, S_y^{s_2 s_3} (l) \, M_{xy}^{\{s\}} (l)\\
 & \quad + \sum_{\{s\}} \int \! d'l \,  S_z^{s_1 s_4} (l) \, S_z^{s_2 s_3} (l) \, M_z^{\{s\}} (l) \, ,
 \end{aligned}
\end{equation*}
 where $ D^{\{s\}} $, $ N^{\{s\}} $, $ M_{xy}^{\{s\}} $, and $ M_z^{\{s\}} $ are exchange propagators that still depend on four Nambu indices.
 They account for the renormalization of the bare interaction $ \mathcal{A}^{(4)} $ in the Cooper, CDW, and in-plane and along-axis magnetic channel, respectively.

 Following the above guiding principle, we then drop interaction terms that conserve momentum only up to $\mathbf{Q}$. In addition, we only take bilinears around the important
 ordering momenta into account, which are $\mathbf{0}$ in the Cooper and $\mathbf{Q}$ in the particle-hole channels. 
 More formally, we may account for the Nambu-index dependence by using the $2 \times 2$ unit matrix $\tau^0$ and Pauli matrices $ \tau^i $, where
 $i = 1,2,3 $. (In contrast, we denote the Pauli matrices by $\tau^x$, $\tau^y$, and $\tau^z$ when they are used for the spin-index dependence.)
  In this notation, we have
\begin{align*}
D^{\{ s\}}(l) & \approx \tau_{s_1 s_2}^0 \tau_{s_3 s_4}^0 \, D (l) \, , \\
  M_{xy}^{\{ s\}}(l) & \approx \tau_{s_1 s_4}^1 \, \tau_{s_2 s_3}^1 \, M_{xy} \, ,\\
  M_z^{\{ s\}}(l) & \approx \tau_{s_1 s_4}^1 \, \tau_{s_2 s_3}^1 \, {M_z} (l) \, ,\\
  N^{\{ s\}}(l) & \approx \tau_{s_1 s_4}^1 \, \tau_{s_2 s_3}^1 \, N (l)
\end{align*}
 with Nambu-index-independent exchange propagators $D (l) $, $M_{xy} (l)$, $M_z (l)$, and $N (l)$.
 Again note that the attribution of bilinears with $ s_1=s_2$ and $s_1=-s_2$ to ordering momenta $\mathbf{l}$ and $ \mathbf{l} + \mathbf{Q}$ only holds
 in the sense of a gradient expansion.
 The approximations made here are fully compatible with the Pauli principle and the particle-hole symmetry
 protecting perfect nesting (for a more detailed discussion, see
 Appendix~\ref{sec:AFM-symm}). 
 The pseudospin SU(2) symmetry couples the CDW and Cooper channels and therefore $ D(l) = - N(l) $. 

 For consistency reasons, we also drop all normal (i.e., time-reversal invariant and momentum-conserving)
 contributions to the self-energy and only the anomalous
 self-energy [i.e., the gap $\Delta (k)$] flows.
 The quadratic part of the action or, in other words, the inverse of the full propagator, reads as
\begin{equation*}
 \begin{aligned}
 \mathbf{C}_\uparrow (k) &= \Delta (k) \, \tau^1 + i k_0 - \epsilon_\mathrm{a} (\mathbf{k}) \, \tau^3 \, ,\\
 \mathbf{C}_\downarrow (k) &= - \Delta^\ast (k) \, \tau^1 + i k_0  - \epsilon_\mathrm{a} (\mathbf{k}) \, \tau^3 \, ,
 \end{aligned}
\end{equation*}
 where $\epsilon_\mathrm{a} (\mathbf{k}) $ denotes the bare dispersion.

  The corresponding fRG flow equations are derived in Appendix~\ref{sec:deriv-ex-para}, and we only give the main result here.
 Due to symmetries, the exchange propagators and the gap are real-valued and perfect nesting implies $ D(l) = - N(l) $ due to the resulting particle-hole symmetry.
 In these final flow equations,
 the following fermionic loops appear:
\begin{align*}
 I_\mathrm{eq}^{\{s'\}} (l,p) & = \frac{1}{2} \left[
 G_\uparrow^{s'_1 s'_2} (p-l/2) \, G_\uparrow^{s'_3 s'_4} (p+l/2) \right. \\ & \qquad  \left.
+ G_\downarrow^{s'_1 s'_2} (p-l/2) \, G_\downarrow^{s'_3 s'_4} (p+l/2) \right]\, ,
\end{align*}
\begin{align*}
 I_\mathrm{op}^{\{s'\}} (l,p) & = \frac{1}{2} \left[
 G_\uparrow^{s'_1 s'_2} (p-l/2) \, G_\downarrow^{s'_3 s'_4} (p+l/2)  \right. \\ & \qquad  \left.
+ G_\downarrow^{s'_1 s'_2} (p-l/2) \, G_\uparrow^{s'_3 s'_4} (p+l/2) \right]\, .
\end{align*}
 They enter via the combinations
\begin{align*}
 I_\mathrm{eq} (l,p) = \frac{1}{4} & \left[
 I_\mathrm{eq}^{++--} (l,p) + I_\mathrm{eq}^{--++} (l,p)
\right. \\ & \quad \left.
 + 2 I_\mathrm{eq}^{+-+-} (l,p) \right] \, , \\
 I_\mathrm{op} (l,p) = \frac{1}{4} & \left[
 I_\mathrm{op}^{++--} (l,p) + I_\mathrm{op}^{--++} (l,p)
\right. \\ & \quad \left.
 + 2 I_\mathrm{op}^{+-+-} (l,p) \right] \, .
\end{align*}
 
The exchange propagators then flow according to
 \begin{equation} \label{eqn:red-D}
  \dot{D}  (l)  =  \int \! d'p \, \dot{I}_\mathrm{eq}(l,p) \, \left\{ F [D,(M_z+D)/2 +M_{xy}] (l,p) \right\}^2 \, ,
 \end{equation}
 \begin{align} \notag
 \dot{M}_{xy} (l)  = - & \int \! d'p  \, \dot{I}_\mathrm{op} (l,p) \\ \label{eqn:red-mxy} & \times \left\{ F [M_{xy},M_z/2 +3D/2] (l,p) \right\}^2 \, ,
 \end{align}
 \begin{align} \notag
 \dot{M}_z (l) = - & \int \! d'p \, \dot{I}_\mathrm{eq} (l,p) \\ \label{eqn:red-mz} & \times \left\{ F [M_z, - M_z/2 + M_{xy} +3D/2] (l,p) \right\}^2 \, .
 \end{align}
 In these equations,
 a dot represents a derivative with respect to the scale $\lambda$,
 and, for a Hubbard-type bare interaction of strength $U$, the exchange propagators on the right-hand side enter via
\begin{equation*}
 F [P_1,P_2] (l,p) = 2 U + 2 P_1 (l) + P_2(p) \, .
\end{equation*}
 Note that these flow equations are not restricted to a specific cutoff scheme and that the regulator can be chosen freely.
 
% discuss these flow equations
 The self-energy flows according to
 \begin{align} \notag
 \dot{\Delta} (k)  = - & \int \! d'p \,  S_\uparrow^{+-} (p) \\ \label{eqn:red-se} & \times E [M_z,(D-M_z)/2 + M_{xy},D] (k,p) \, ,
 \end{align}
 where
\begin{equation*}
  E[P_1,P_2,P_3](k,p)  =2 U + 2 P_1 (0) + P_2 (k-p) +P_3 (k+p) \, .
\end{equation*}

 These flow equations have a very similar structure to those in Refs.~\onlinecite{HGS-freq-dep,giering-new,eberlein-sssu}.
 They are complemented by the WI
\begin{align} \notag
  \Delta (k) -\Delta_0 = - & 2 \Delta_0 \int \! d'p\, I_\mathrm{op} (0,p) \\ \label{eqn:red-WI} & \times E [M_{xy}, (M_z+D)/2, D] (k,p) \, .
\end{align}

 Eqs.~(\ref{eqn:red-D})--(\ref{eqn:red-se}) describe the fRG flow in a simple approximation beyond mean-field theory. Corrections to the mean-field picture 
 enter in vertex-correction and box diagrams for the interaction and Fock-type diagrams for the self-energy.
 At the RPA level, such vertex-correction or box diagrams are neglected and the
 formal solution of the flow equations fulfills the WI exactly. Without invoking further approximations, the mean-field gap equation is recovered.

 Beyond RPA, the flow equations have to be solved numerically.
 The Ward identity may then be violated due to the one-loop truncation and due to the approximations underlying the parametrization employed.
 The violation of the WI may therefore be regarded as a measure of truncation and/or parametrization errors.
 
\subsection{Random phase approximation} \label{sec:rpa}

 Let us now consider the flow equations~(\ref{eqn:red-D})--(\ref{eqn:red-mz}) at the RPA level, i.e., neglect $P_2$. These flow 
 equations then take on the form
\begin{equation} \label{eqn:ext-RPA}
 \dot{{P}} (l) = - {P} (l) \, \dot{{B}}_{P} (l)\, {P} (l)  
\end{equation}
 with ${B}_{P} (l) = 4  \int \! d'p \, {I}_P (l,p) $ and $ P(l) = U +  P_1 (l)$.
 Here, $P_1$ may be $D$, $M_z$, or $M_{xy}$ and the loop functions are $ I_\mathrm{op}$ for $M_{xy}$ and $ \pm I_\mathrm{eq}$ for $M_z$ or $D$.
 One can clearly see that these exchange propagators only couple via the self-energy at the RPA level.
 The generic RPA flow equation~(\ref{eqn:ext-RPA}) is solved by
\begin{equation} \label{eqn:form-sol}
 {P}(l) = U \left[ {1} + U {B}_P (l) \right]^{-1} \, ,
\end{equation}
 and this formal solution also fulfills the Bethe-Salpether equation
\begin{equation*}
 {P}(l) = U \left[ {1} - {B}_P (l) \, {P} (l) \right] \, .
\end{equation*}
 
 Let us also neglect self-energy diagrams with bosonic lines inside the loops, i.e., $P_2$ and $P_3$ are sent to zero in
 $ E$ in Eq.~(\ref{eqn:red-se}).
 The resulting approximate flow equation for the self-energy reads as
\begin{equation} \label{eqn:ex-rpa-se}
 \dot{\Delta} = - 2 {M}_z (0) \int \! d'p \; S_\uparrow^{+-} (p) \, .
\end{equation}
 Note that the self-energy loses its momentum and frequency dependence at the RPA level.
 %Furthermore, $\Delta$ and $\Sigma_\mathrm{s} $ must be real due to the frequency inversion symmetry.
 We observe that
\begin{equation*}
 \int \! d'p \; S_\uparrow^{+-} (p) = \int \! d'p \; \dot{G}_\uparrow^{+-} (p) - \frac{1}{2} {B}_\mathrm{eq} (0) \, \dot{{\Delta}}  \, .
\end{equation*}
 By virtue of this identity, inserting the formal solution~(\ref{eqn:form-sol}) for $M_z$ into Eq.~(\ref{eqn:ex-rpa-se}) and integrating yields the 
 mean-field gap equation
\begin{equation} \label{eqn:ext-gap}
  \Delta - \Delta_0 = \Delta \, U \int \! d'\mathbf{k} \, \frac{1}{ \sqrt{\epsilon_\mathrm{a} (\mathbf{k})^2 + \Delta^2} }  \, .
\end{equation}
 
 Let us now discuss the formal solution of the flow equations in RPA and then elaborate on the fulfillment of the WI.
 From the relation 
\begin{equation*}
 \int \! d'k \, G_\uparrow^{+-} (k) = - \frac{1}{2} \Delta B_\mathrm{op} (0) 
\end{equation*} 
 and the gap equation~(\ref{eqn:ext-gap}) one obtains
\begin{equation} \label{eqn:d-d0}
 1 +  U B_\mathrm{op} (0) = \frac{\Delta_0}{\Delta} \, ,
\end{equation}
 and consequently
$ {M}_{xy} (0) = U \left(\Delta / \Delta_0 - 1 \right) $.
 This reflects the Goldstone-vertex nature of $M_{xy} (0) $.
 Neglecting bosonic lines inside closed loops leads to
\begin{equation*}
 \Delta - \Delta_0  = - \Delta_0 \, B_\mathrm{op} (0) \, \left[ U + M_{xy} (0) \right]
\end{equation*}
 for the WI. 
 By inserting the exact solution~(\ref{eqn:form-sol}) for the Goldstone vertex, Eq.~(\ref{eqn:d-d0}) is reproduced, and hence
 the RPA solution is fully consistent with the WI.
As in Ref.~\onlinecite{eberlein-sssu} for a singlet superconductor, one may write in leading order in a gradient expansion
 with coefficients $\alpha $ and $\beta$
\begin{equation*}
 M_{xy} (l) \propto \frac{1}{\Delta_0 + \alpha l_0^2 + \beta \mathbf{l}^2}
\end{equation*}
 in the limit $\Delta_0 \to 0$. Beyond RPA, this property appears likely to be preserved by the WI~(\ref{eqn:red-WI}), 
 provided that $M_{xy} $ remains the only propagator which diverges for a vanishing seed field.

\section{Application to a simple two-pocket model}  \label{sec:appl}

\subsection{Model} \label{sec:chubu}
 
\begin{figure}
 \centering
\includegraphics[width=.8\linewidth]{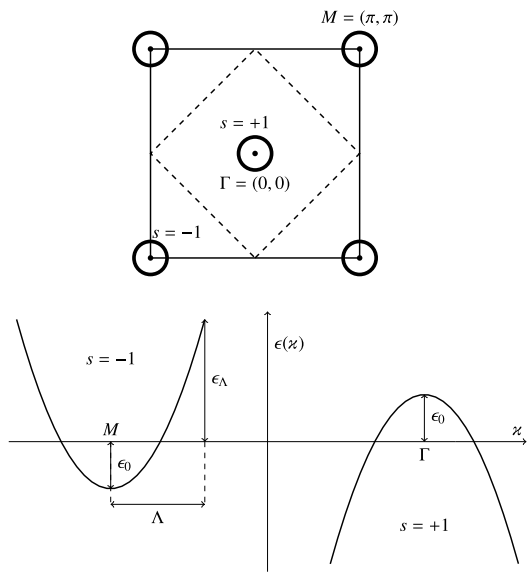}
 \caption{Dispersion of the two-pocket model. Upper panel: Fermi surfaces (bold lines) in the full BZ for the physical momentum
 $ \varkappa = \mathbf{k} + (1-s)/2 \, (\pi,\pi) $. The dashed line represents the boundary of the \emph{reduced} BZ, on which the momentum quantum number
 $\mathbf{k}$ is defined. Lower panel: Dispersion as a function of the physical momentum $\varkappa$ along the diagonal of the full BZ.
 The dispersion is cut off at the energy $ \epsilon_\Lambda= \Lambda^2 /2 - \epsilon_0$. See text for further explanation.}
 \label{fig:disp-chubu}
\end{figure}

 In this Section, we numerically integrate
 the flow equations (\ref{eqn:red-D})--(\ref{eqn:red-se}) as a first step beyond the mean-field picture for the AF phase within a fRG framework.
 In the derivation of these equations, a perfectly nested dispersion has been assumed.
 As an example of such a model, the repulsive Hubbard model in two dimensions with hopping only between nearest neighbors
 has already been mentioned 
 above. Studying its flow into the antiferromagnetic phase would complement recent work on the superfluid phase. \cite{eberlein-repu,eberlein-phd}
 In order to get some intuition, it seems, however, preferable to consider a model with a higher symmetry, which will require less computational
 resources.
 Good candidates for such a model are effective low-energy theories, e.g., (extended) $g$-ology models.
 In this work, a two-pocket model \cite{chubukov} in two dimensions proposed by Chubukov \emph{et al.\ }will be considered. Originally, it was conceived for a (Wilsonian) RG study of the
 competition between spin-density-wave (SDW) order and superconductivity in the iron pnictides. Having a purely quadratic
dispersion and a simplified momentum dependence of the interactions, this model has an
 ultraviolet cutoff $\Lambda$. The remaining degrees of freedom live on two patches centered around the $ \Gamma$ and the $M$ points in the folded
 two-dimensional Brillouin zone (BZ) and mimic the band structure close to the Fermi surfaces around these points.
 (For a pictorial representation of the dispersion, see Fig.~\ref{fig:disp-chubu}.)
 In the following, this folded BZ
 will be referred to as the \emph{full} BZ in order to avoid confusion with the \emph{magnetic} BZ, which is bounded by the dashed line in the upper panel
 of Fig.~\ref{fig:disp-chubu}.

 The bare action of the two-pocket model will now be expressed in terms of the Nambu spinors $\Psi_\sigma (k)$ in Eq.~(\ref{eqn:Nambu-spinor}) with components
 $\Psi_\sigma^s (k)$, where the subscript $\sigma$ denotes the spin projection.
 In this case, the Nambu indices $s$ can as well be interpreted as pocket indices,
 where $s=+1$ and $s=-1$ correspond to the hole pocket at the $\Gamma $ point and the electron pocket at the $M$ point, respectively. 
 The momentum quantum numbers $\mathbf{k}$ therefore vary only within the pockets (see also Fig.~\ref{fig:disp-chubu}.)
 In Nambu space, the bare action then reads as
\begin{equation*}
 \mathcal{A} = \sum_\sigma \int_{|\mathbf{k}| \leq \Lambda} \!\!\! d k \, \bar{\Psi}_\sigma (k) \, C_\sigma (k) \, \Psi_\sigma (k) + \mathcal{A}^{(4)} [\bar{\Psi}, \Psi] \, ,
\end{equation*}
where $ C_\sigma (k) $ is of the form given in Eq.~(\ref{eqn:Cform}) with
\begin{equation*}
  \epsilon_\mathrm{a} (k) = - \frac{\mathbf{k}^2}{2} + \epsilon_0 \quad \text{and} \quad \epsilon_\mathrm{s} =0 \, .
\end{equation*}
 In the one-particle dispersion, we have set the fermionic mass to unity and we will use natural units in the following, i.e.\ energies appear as dimensionless quantities.
 Note that, in two dimensions, this dispersion corresponds to a constant density of states $\rho_0 = 1 /(2 \pi) $.
 In the following, $\epsilon_0 > 0$ so that there are two circular Fermi surfaces centered around the $M$ and the $\Gamma$ points.

The bare interaction reads as 
\begin{align*}
 \mathcal{A}^{(4)} & = - \sum_{s_1 \dots s_4} \sum_{\sigma_1 \dots \sigma_4} \int _{|\mathbf{k}_i| \leq \Lambda} \!\!\! d k_1 \dots  d k_4 \bar{\Psi}^{s_1}_{\sigma_1} (k_1) \, \bar{\Psi}^{s_2}_{\sigma_2} (k_2) \\
& \quad \times {\Psi}^{s_3}_{\sigma_3} (k_3) \,{\Psi}^{s_4}_{\sigma_4} (k_4)\, \delta_{\{k\}} \delta_{\sigma_1 \sigma_4} \, \delta_{\sigma_2,\sigma_3} \, \\
 & \quad \times \left\{ U_1 \, \delta_{\mathbf{s},(+--+)} + U_2 \, \delta_{\mathbf{s},(-+-+)} \right. \\ & \quad \times \left.
 + \frac{U_3}{2} \left[ \delta_{\mathbf{s},(--++)}  +\delta_{\mathbf{s},(++--)} \right]  \right. \\
 & \qquad \left. + \frac{U_4}{2} \left[ \delta_{\mathbf{s},(++++)}  +\delta_{\mathbf{s},(----)} \right] \right\}
\end{align*}
with bare couplings $U_i $. Both quartic and quadratic parts respect the
 positivity and the particle-hole symmetry discussed in Appendix~\ref{sec:AFM-symm} and hence the fRG flow equations preserve these symmetries.
 In the following, only the case $ U_1 = U_2 = U_3 = U_4 = U $ is studied.
 The interaction then has the same form as a Hubbard interaction expressed in momentum
space. The only difference to the Hubbard model in the two-patch approximation then lies in the dispersion, which is isotropic in the present case.
  (Note that flipping a Nambu index corresponds to a momentum shift by $\mathbf{Q}$.)

 In order to study the flow into the SDW phase, a small symmetry-breaking seed field $ \Delta (k) =\Delta_0 $  is added to the bare action.
 This will regularize divergences resulting from the Goldstone modes.
 The case of the \emph{spontaneous} breaking of the SU(2) and translational symmetries is recovered in the limit $ \Delta_0 \to 0 $. In practice, this means
 that $\Delta_0$ is chosen to be small compared to the other energy scales in the bare action.
 After the infrared cutoff $\lambda $
has been removed by the RG flow, $\Delta_0$ may subsequently used as a flow parameter which is sent to zero. \cite{eberlein-sssu,eberlein-phd} However, we will refrain from considering a
 seed-field flow, since the focus in this work rather lies on more basic questions such as the applicability of our approximate parametrization. 

 Also in the presence of such a seed staggered magnetization $\Delta_0$, the two-pocket model is momentum-conserving in the basis of pseudospinors
 defined in Eq.~(\ref{eqn:pseudospin}). This implies that
 $ \operatorname{Re} \epsilon_\mathrm{s} $ and $\operatorname{Im} \epsilon_\mathrm{a}$ vanish at all instances of the RG flow. (This can be more easily shown if one chooses the 
 physical spins to align in the $x$-  instead of the $z$-direction.) Moreover, this hidden symmetry only allows for non-momentum-conserving interactions if these terms
 also break the time-reversal symmetry. In a way, this \emph{a posteriori} justifies the simultaneous omission of interactions breaking at least one of these discrete
 symmetries in Sec.~\ref{sec:param-gen}.

 From a numerical viewpoint, it is preferable to choose this low-energy continuum model instead of a lattice model for a first fRG study of the AF phase beyond mean field.
 First,  such an effective model may allow for a parametrization of its renormalized coupling functions based on a gradient expansion.
 In addition, the $ C_{4v} $ symmetry of a 2D lattice model such as the ones used in Refs.~\onlinecite{wang-2009,wang-2010,Thomale-Platt-pnictides,platt-first,pnictides-gap,sidplatt} is promoted to a full circular symmetry, which imposes
 more severe restrictions on the allowed terms in such a gradient expansion and simplifies the integration over internal momenta in Feynman diagrams.
 Altogether, this will lead to a considerable reduction of the numerical effort undertaken in a numerical integration of the flow equations.

In Ref.~\onlinecite{chubukov}, the RG flow of this model has been analyzed in the symmetric phase with momentum- and frequency-independent couplings $U_1,U_2,U_3 $, and $U_4$. This can be regarded as a gradient-expansion approach in leading order. Obviously, one has to go beyond this approximation in the symmetry-broken phase, since the 
violation of the SU(2) Ward identity (\ref{eqn:red-WI}) would otherwise be horrendous.
  For the case $ U_1 = U_2 = U_3 = U_4 \equiv U $ studied here, the mean-field gap equation
\begin{equation} \label{eqn:MF-chubu}
 \Delta =  2U \int \! d'k\, \frac{\Delta}{k_0^2+ \epsilon_\mathrm{a} (\mathbf{k})^2 + \Delta^2}
\end{equation}
 for antiferromagnetism has the same form as for the Hubbard model at half-filling (see, for example, Ref.~\onlinecite{Onsager-Hubbard}) and the BCS gap equation.
 For the AF case, the prime in the measure $d'k$ indicates that the corresponding momentum integral only runs over half the BZ.
 Since the two-pocket model has a constant density of states $ \rho = \rho_0 \equiv 1 /(2\pi )$ between $ \epsilon_\mathrm{a} = - \epsilon_0 $ and $ \epsilon_\mathrm{a} = \epsilon_\Lambda
 \equiv \Lambda^2/ 2 - \epsilon_0 $ and $ \rho =0$ outside this low-energy window, the momentum integral can be performed analytically. This yields
\begin{equation*} 
 \rho_0 U \left[ \mathrm{Arsinh} \left( \frac{\epsilon_0}{\Delta} \right) + \mathrm{Arsinh} \left( \frac{\epsilon_\Lambda}{\Delta} \right) \right] = 1\, .
\end{equation*}
 Clearly, there is no critical interaction strength, i.e., for any positive value of $U$ there will be a finite gap.
 
 As an approximate solution at weak coupling $  U \ll 1/\rho_0 $, we have
\begin{equation*}
 \Delta \approx 2 \sqrt{\epsilon_\Lambda \epsilon_0} \exp \left( - \frac{1}{2 \rho_0 U} \right) \, .
\end{equation*}
 
 Note that the two-pocket model is not safe against a variation of the ultraviolet cutoff, as the gap grows with $ \sqrt{\epsilon_\Lambda} $
 at weak coupling.
 %This might be a shortcoming of mean-field theory. If such a behavior can also be observed in the fRG results (or the results obtained from
 %another low-energy solver), this would contradict the concept of an effective low-energy theory, according to which low-energy observables
 %should only vary weakly with the cutoff.
However, since our focus rather lies on the methodology than on real materials, this lack of UV safety does not really pose a problem.
 
\subsection{Numerical implementation} \label{sec:chubu-impl}
 Let us now turn to the implementation of the fRG flow equations (\ref{eqn:red-D})--(\ref{eqn:red-se}) for the two-pocket model.  
 The circular symmetry of this model will be exploited and all calculations will be performed at zero temperature.

 For the low-energy model considered, it seems appealing to parametrize the momentum dependence instead of resorting to a discretization 
 in momentum space.
 This will considerably lower the numerical effort spent on the integration of the flow equations~(\ref{eqn:red-D})--(\ref{eqn:red-se}).
 In order to keep the momentum dependence simple, a frequency cutoff seems preferable to other schemes.
 An additive frequency regulator will turn out to be a good choice in the following. As in Refs.~\onlinecite{eberlein-sssu,eberlein-repu,eberlein-phd},
 the infrared cutoff $\lambda$ is implemented by the replacement
\begin{equation} \label{eqn:cutoff}
 i k_0 \to i k_0 + R_\lambda (k_0) = i \, \mathrm{sign} (k_0) \, \sqrt{k_0^2 + \lambda^2}
\end{equation}
 in the quadratic part of the bare action. One might also consider a multiplicative regulator as 
 in the $\Omega$ scheme of Refs.~\onlinecite{husemann_09,HGS-freq-dep,giering-new}, but that cutoff scheme would lead to more demanding loop integrals on the
 right-hand sides of the fRG flow equations.

 Note that the frequency dependence of the vertices and the self-energy may not be easy to parametrize at finite scales.
 In the following, a parametrization of the momentum dependence is given, where the coefficients all remain frequency dependent.
 This latter dependence is then discretized using a logarithmic grid.

 Let us first address the momentum dependence of the exchange propagators.
 In the spirit of a gradient expansion around ordering momenta, one may approximate the momentum dependence of each exchange propagator $ P (l) $ by a 
 Lorentzian, i.e.,
\begin{equation} \label{eqn:gradient-exp}
  P (l) = \frac{1}{m_P (l_0) \left[1+n_P (l_0) \, \mathbf{l}^2 \right]} \, ,
\end{equation}
 with two frequency-dependent parameters.
 $ m_P (l_0) $ corresponds to a bosonic mass and determines the height of the Lorentz peak at $ \mathbf{l} = 0 $ with width $ \left|n_P (l_0) \right|^{-1/2} $. 
 In practice, $ n_P$ will be determined from a finite difference formula for $ 1/P(l)$.

 Due to the (continuous) rotation symmetry of the model, corrections to 
 this ansatz would appear as even-order terms in $|\mathbf{l}| $ in the denominator.  
 %Note that this ansatz may also describe precursors of Landau damping. \textsc{Explain this in further detail}
 Conversely,
 a frequency-dependent \mbox{$g$-ology} approach
 would correspond to
 neglecting the $ \mathbf{l}^2$ term in the denominator.
 In a mixed fermion-boson fRG approach to superfluidity in the attractive Hubbard model, however,
 gradient terms of radial and Goldstone modes are reminiscent of the above Lorentz decay. \cite{strack-superfluid,obert-superfluid}
 Therefore, it seems prudent to at least include the $\mathbf{l}^2$ term in Eq.~(\ref{eqn:gradient-exp}).
 In order to keep the computational cost low, we will restrict ourselves to this lowest nontrivial order in a first attempt of a fRG study in the AF phase beyond the
 mean-field picture.
% But in the implementation of the RG flow equation used here, the inclusion of higher-order terms
% does not create parallelization issues in principle.

 In the following, the gap functions will be projected to zero momentum, i.e., we work with a frequency-dependent gap
\begin{equation*}
\Delta (k_0) = \left. \Delta (k) \right|_{\mathbf{k}=0} \, .
\end{equation*}
 Note that $\mathbf{k}=0$ corresponds to considering the gap only at the centers of the pockets.
 The Nambu indices play the role of pocket indices,
 $\mathbf{k}$ therefore lives on half the BZ (the magnetic BZ) and only varies \emph{within} the pockets.
 Of course, also the momentum dependence of the gap would be interesting to study and resolving only
 its frequency dependence may seem sloppy at first. Looking at the flow equation~(\ref{eqn:red-se}) for the self-energy, one can, however, observe that
 the frequency and momentum dependence is generated by the dependence of the second and third arguments in the square brackets of $E$.
 Since only terms up to order $\mathbf{l}^2$ in a gradient expansion
 are contained in our parametrization of the exchange propagators
 and since the self-energy is mainly driven by the
 radial vertex at $l=0$, it seems appropriate to neglect the momentum dependence of the self-energy in 
 a first step beyond MFT. This way, the integrand in Eq.~(\ref{eqn:red-se}) remains independent of the angular integration variable,
  which reduces the three-dimensional integral to a two-dimensional one.
 Studying the momentum dependence of the self-energy appears, however, worthwhile if one goes beyond a Lorentzian profile in the exchange propagators.

 In the following, numerical results for the fRG flow into the SDW phase of the two-pocket model are presented.
 The system parameters are chosen as $ \epsilon_0 = 3.0 \cdot 10^{-2} $ and $ \epsilon_\Lambda = 0.58 $, if not indicated otherwise.
 All calculations are performed at zero temperature.
 \subsection{Scale dependence of the exchange propagators and the gap} \label{sec:numflow}
\begin{figure}
  \centering
  \includegraphics[width=\linewidth]{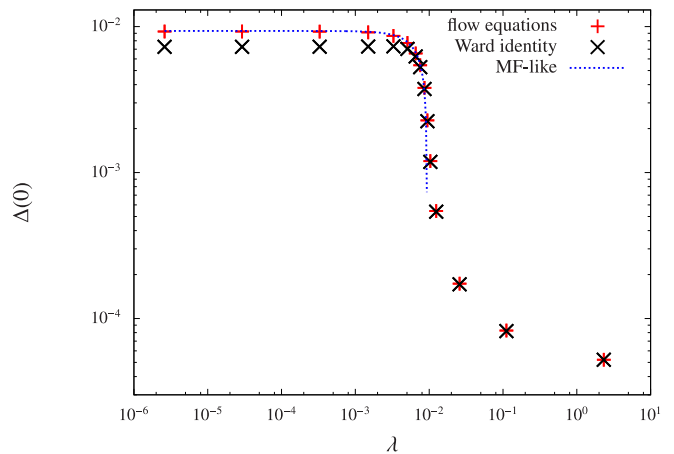}
  \caption{(Color online) Flow of the gap $ \Delta (0) $  with the infrared cutoff $\lambda$ at zero frequency for $U=1.0$ and $ \Delta_0 = 5.0 \cdot 10^{-5} $.
  The points labeled with `flow equation' are obtained from the integration of Eqs.~(\ref{eqn:red-D})--(\ref{eqn:red-se}), while the points labeled with
 `Ward identity' correspond to the value of the right-hand side of the WI~(\ref{eqn:red-WI}) at the respective scale. The curve labeled as
 `MF-like' corresponds to a fit of the former data set to $ \Delta_\lambda (0) = \sqrt{\alpha^2 - \lambda^2} $ below the critical scale
  with the fit parameter $\alpha$. See text for further explanation.}
  \label{fig:delta-flow}
 \end{figure}

\begin{figure*}
  \centering
  \includegraphics[width=\linewidth]{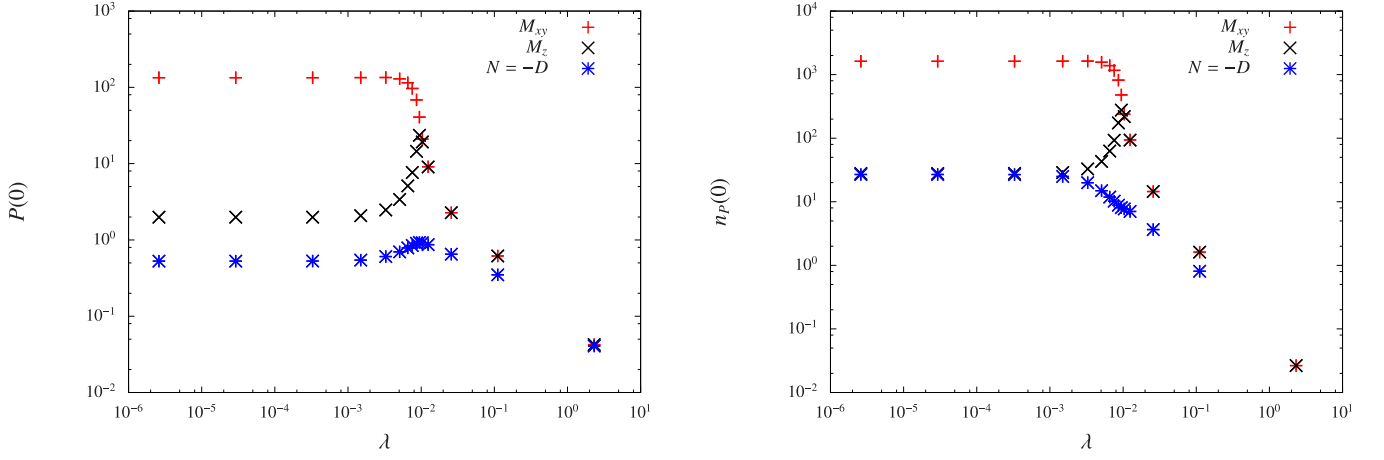}
  \caption{(Color online) Flow of the parameters of the exchange propagators with the infrared cutoff $\lambda$ at $l=0$ for $U=1.0$ and $ \Delta_0 = 5.0 \cdot 10^{-5} $.}
  \label{fig:prop-flow}
 \end{figure*}

 Let us first consider the flow of the gap and the exchange propagators at zero momentum and frequency.
  For a typical choice of the model parameters, the scale-dependence of these quantities is depicted in Figs.~\ref{fig:delta-flow}
 and \ref{fig:prop-flow}, respectively. In qualitative agreement with the mean-field picture (cf.\ Ref.~\onlinecite{me-AFM-MF}),
 the gap opens at the critical scale, where the radial vertex $M_z (0) $ shows a pronounced peak. 
 Below the critical scale, the couplings saturate to their infrared values. In contrast to the radial vertex, which has moderate infrared values,
 the Goldstone vertex $M_{xy} (0)$ becomes large for $\lambda \to 0$.

 Below the critical scale, the flow of the gap behaves mean-field like. For the cutoff chosen in Eq.~(\ref{eqn:cutoff}) and in MFT, the scale dependence of the
 gap takes on
 the form $ \Delta_\lambda = \sqrt{ \Delta_{\lambda=0}^2 - \lambda^2} $. Indeed, that scale dependence can also be observed for the fRG gap below the critical scale
 in Fig.~\ref{fig:delta-flow}. We hence conclude that the coupling between different interaction channels below the critical scale has a negligible impact
 on the infrared value of the gap.

 If the seed field is varied,
 one finds that the increase of the gap is steeper for smaller $\Delta_0$ and that the peak of the
 radial vertex is then more pronounced. Moreover, the infrared value of the Goldstone vertex increases with decreasing $\Delta_0$. This behavior
 is also in qualitative agreement with the mean-field results of Ref.~\onlinecite{me-AFM-MF}, while the mean-field picture becomes inadequate on a more quantitative
 level. One difference shall already be outlined here: While the CDW and singlet-pairing channels do not feed back on the other channels and the gap at the
 mean-field level, they will be found to have a significant impact on the flow in Sec.~\ref{sec:comp-MF}. 

At $l=0$, the corresponding exchange
 propagators $ N $ and $D=-N$ grow in the flow until the critical scale is reached. Below, they decrease slightly, saturating to their infrared values.
 Their absolute values are equal due to the pseudospin SU(2) symmetry discussed in Appendix~\ref{sec:AFM-symm}
 and they virtually behave independently of the value of $\Delta_0$.

 After having discussed the flow of the exchange propagators at $l=0$,
 let us now turn to their dependence on momentum and frequency.
 In the right panel of Fig.~\ref{fig:prop-flow}, the flow of the squares $n_P$ of the inverse Lorentz widths of their momentum profiles is depicted at $ l_0 =0$. 
 (Large values of $n_P$ correspond to narrow peaks.)
 Comparison with the values of $m_P$ on the left panel of Fig.~\ref{fig:prop-flow} suggests as a rule of thumb that the momentum profile of the exchange propagators around $l=0$
 is the more sharply peaked the larger their values at $l=0$ are.

\begin{figure*}
  \centering
  \includegraphics[width=\linewidth]{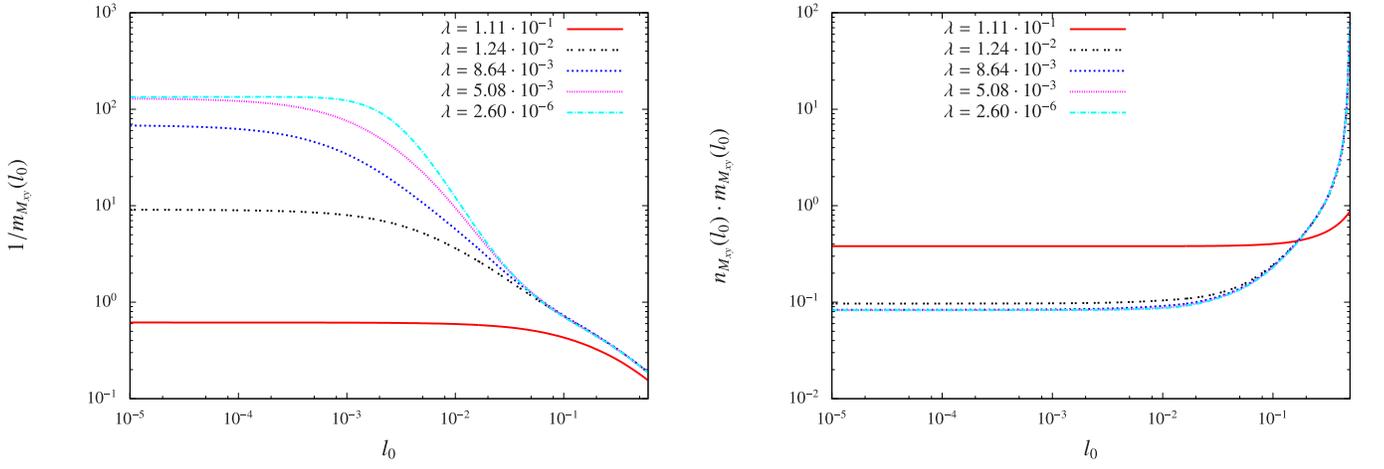}
\caption{(Color online) Frequency dependence (left) of $M_{xy} $ at $\mathbf{l}=0$ and (right) of the Lorentz width for its momentum decay 
 for $U=1.0$ and $ \Delta_0 = 5.0 \cdot 10^{-5} $ at various stages of the flow, where $l_0$ denotes the transfer frequency. The curves shown here
 are the spline interpolants also used in the numerics.}
  \label{fig:gol-freq}
\end{figure*}

\begin{figure*}
  \centering
  \includegraphics[width=\linewidth]{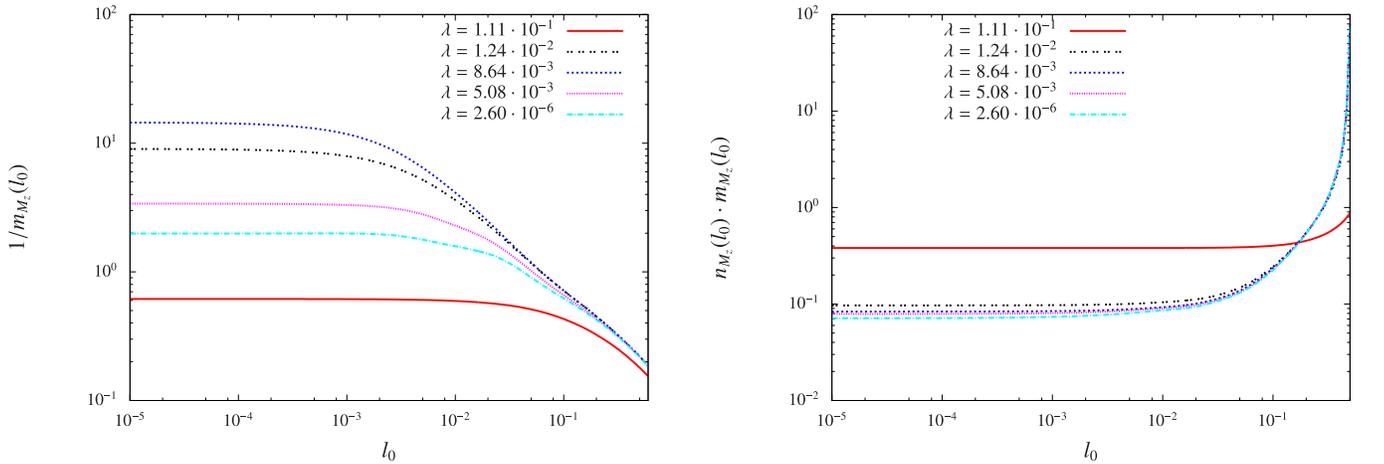}
\caption{(Color online) Frequency dependence (left) of $M_z $ at $\mathbf{l}=0$ and (right) of the Lorentz width for its momentum decay
 at various stages of the flow, where $l_0$ denotes the transfer frequency. The parameters have been chosen as in Fig.~\ref{fig:gol-freq}.}
  \label{fig:rad-freq}
\end{figure*}

\begin{figure*}
  \centering
  \includegraphics[width=\linewidth]{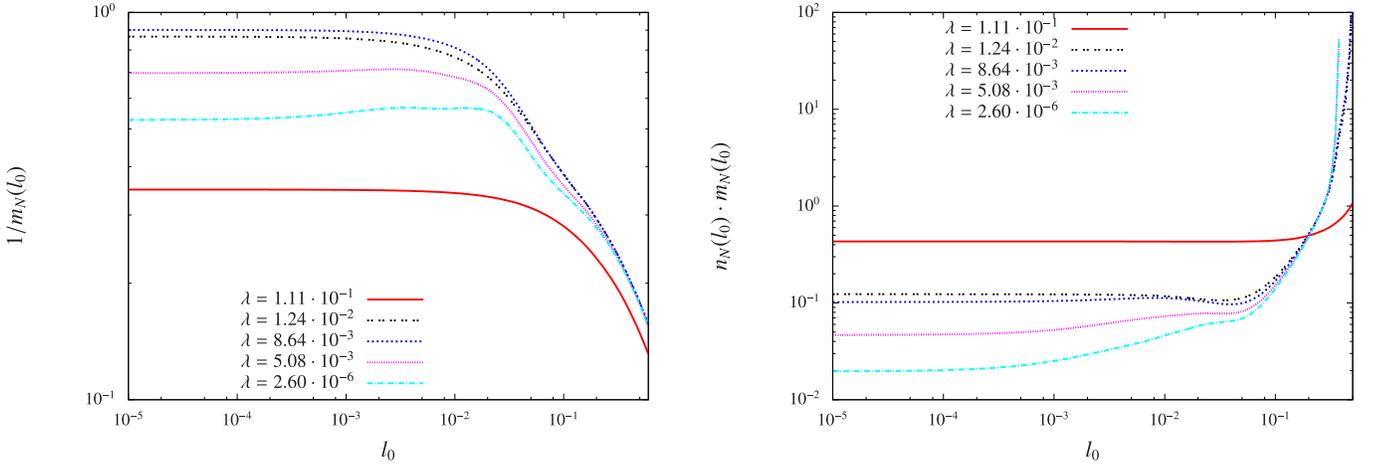}
\caption{(Color online) Frequency dependence (left) of $N $ at $\mathbf{l}=0$ and (right) of the Lorentz width for its momentum decay 
 at various stages of the flow, where $l_0$ denotes the transfer frequency. The parameters have been chosen as in Fig.~\ref{fig:gol-freq}.}
  \label{fig:cdw-freq}
\end{figure*}

 Let us now have a look at the frequency dependence of the exchange propagators $P$. For zero momentum, they are given by $m_P (l_0)$.
 The value of $n_P (l_0)$, in contrast, describes the momentum decay at some frequency $l_0$.
 For $P = M_{xy}, M_z , N$, the parameters $1/m_P $ and $n_P$ are depicted as a function of frequency at various stages of the flow
 in Figs.~\ref{fig:gol-freq}, \ref{fig:rad-freq}, and \ref{fig:cdw-freq}, respectively.
 In the Goldstone and radial channels, $1/m_{M_{xy}} $ and $1/m_{M_z} $ decay monotonically as shown in the left panels of Figs.~\ref{fig:gol-freq} and \ref{fig:rad-freq}.
 The form of these curves neither resembles a Lorentzian nor an exponential.
 One may wonder whether a sign change occurs in $D$ and $N$ in 
 analogy to the superconducting phase of the attractive Hubbard model, where the magnetic exchange propagator changes sign at small frequencies. \cite{eberlein-sssu}
 From Fig.~\ref{fig:cdw-freq}, however, one can see that this is not the case and that $ 1/m_N$ decays with frequency in a way
 similar to the Goldstone and radial channels.
 This is presumably due to the pseudospin SU(2) symmetry [cf.~Eq.~(\ref{eqn:pseudospin})] of the two-pocket model discussed here.

 The parametrization of the exchange propagators introduced in Sec.~\ref{sec:chubu-impl} also allows for a frequency-dependent momentum decay length.
 In the right panels of Figs.~\ref{fig:gol-freq} and \ref{fig:rad-freq}, however,
 the product $ n_P (l_0) \, m_P (l_0) $ for the radial and Goldstone vertices remains constant
 up to relatively high frequencies. This is in agreement with the parametrization of the exchange propagators in Ref.~\onlinecite{eberlein-sssu}.
 In that work, real-valued exchange propagators $P$ are described by a frequency-dependent mass $m_P (l_0)$ and a momentum function $ F_P (\mathbf{l}) $ 
 according to
\begin{equation} \label{eqn:eberlein-momentum}
 P (l) = \frac{1}{m_P (l_0) + F_P (\mathbf{l})} \, .
\end{equation}
 In terms of a gradient expansion in momentum and frequency, such a parametrization applies whenever mixed terms are of minor importance.
 
 In the present case, one may approximate $  F_P (\mathbf{l}) \approx n_P (0) \, m_P (0) \, \mathbf{l}^2  $, which can save half the computation time.
 For more refined momentum parametrizations, however, one may gain a much larger factor by neglecting the frequency dependence of the momentum decay
 in the spirit of Eq.~(\ref{eqn:eberlein-momentum}).
 According to Fig.~\ref{fig:cdw-freq}, these approximations seem less applicable for the CDW and singlet-pairing channels, where
 $ n_N (l_0) \, m_N (l_0) $  varies at frequencies of the order of the critical scale. But, employing Eq.~(\ref{eqn:eberlein-momentum})
 also for these channels should nevertheless affect the results for the gap and the fulfillment of the WI only insignificantly.
 
 In contrast to the exchange propagators, the gap $\Delta$ only shows a negligible frequency dependence throughout the flow in agreement with the RPA result in
 Sec.~\ref{sec:rpa}.
 \subsection{Fulfillment of the Ward identity} \label{sec:WI-fulfill}
 The violation of the WI (\ref{eqn:red-WI}) provides a measure for the errors induced by the Katanin one-loop truncation and all subsequent approximations.
 For the superconducting phase of the attractive Hubbard model,
 the corresponding U(1) WI has been used as a measure of the quality of the approach pursued. \cite{eberlein-sssu}
 In the present case, having a look at the violation of the SU(2) WI seems indeed rewarding, since there are a number of approximations involved
 and since it is not yet clear how faithful they are on a more quantitative level.
 
\begin{figure*}
  \centering
  \includegraphics[width=\linewidth]{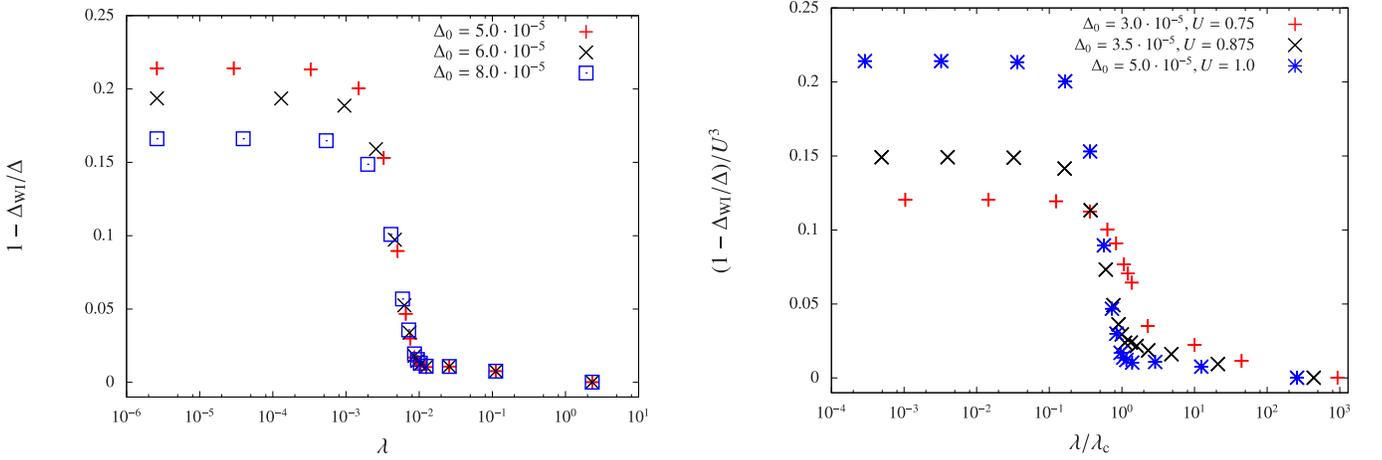}
\caption{(Color online) Relative violation of the WI~(\ref{eqn:red-WI}) as a function of the scale $\lambda$ for various values of $U$ and $\Delta_0$. 
The fRG results for the gap are denoted by $\Delta$, while $\Delta_{\rm WI}$ is obtained from the right-hand side of the WI~(\ref{eqn:red-WI}).
 The interaction strength $U= 1.0$ in the left plot.
 In the right panel, the relative WI violation is rescaled by $U^3$ and  the scale $\lambda$ by the corresponding critical scale $\lambda_{\rm c}$.
} \label{fig:wis-u}
\end{figure*}

 In Fig.~\ref{fig:wis-u}, the relative WI violation $ \left( \Delta - \Delta_{\rm WI} \right) / \Delta $ is plotted against the scale, 
 where $\Delta$ is obtained from 
 the fRG flow equations and $ \Delta_{\rm WI} $ denotes the corresponding value of the right-hand side of Eq.~(\ref{eqn:red-WI}).
 Obviously, perturbation theory applies at high scales  and the WI is only weakly violated in that regime.
 Slightly above the critical scale, the curves in Fig.~\ref{fig:wis-u} start to increase and develop a dependence on the value of the seed field $\Delta_0$.
 Generically, $\Delta$ is larger than $\Delta_{\rm WI}$ and the WI violation gets worse for smaller seed fields. 
 For the parameters of Fig.~\ref{fig:wis-u}, the values of the WI violation ($\leq 25 \%$) suggest that the results obtained have at least
 the right order of magnitude,
 while they are less faithful than in Ref.~\onlinecite{eberlein-sssu}, where the relative WI violation is smaller.
 Regarding the underlying approximations, this suggests that the flow equations~(\ref{eqn:red-D})--(\ref{eqn:red-se}) in
 Nambu-normal approximation are indeed applicable, while this approach should be extended in an attempt to proceed in a more quantitative direction.

 The impact of different approximations on the WI can also be assessed by looking at the dependence of its violation on the interaction strength.
 In the Katanin scheme, the relative WI violation is expected to grow as $U^3$
(cf.\ Refs.~\onlinecite{eberlein-sssu,eberlein-phd}).
 As can be seen in Fig.~\ref{fig:wis-u}, our results do not coincide with this expectation. Instead, we find
considerable contributions to the WI violation that scale as $U^2$, also above the critical scale. This can be regarded as a signature of the
approximations made within the one-loop truncation, for example the projection rule of Appendix~\ref{sec:AFM-ex-para} and the omission of some interaction
terms.
\begin{figure}
  \centering
  \includegraphics[width=\linewidth]{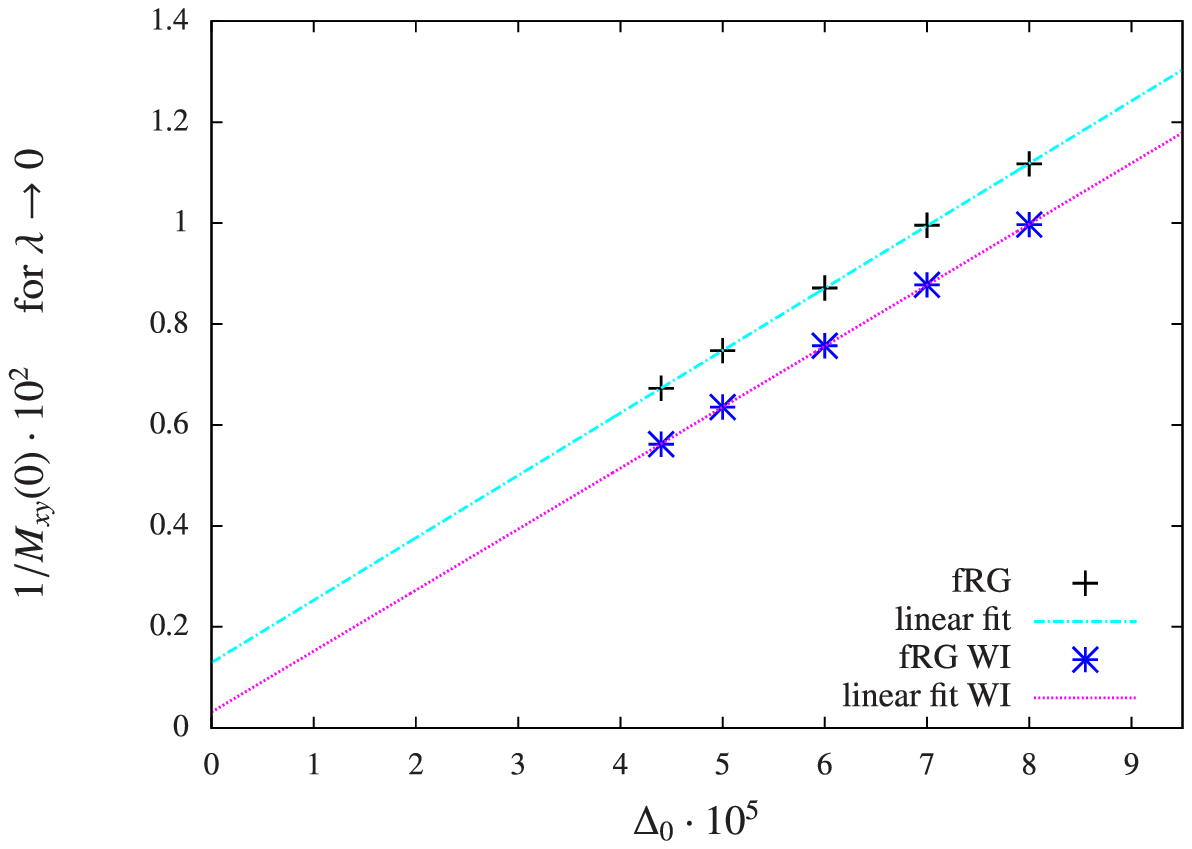}
\caption{(Color online) Dependence of the infrared values of the Goldstone vertex $M_{xy} (0)$ on the seed field $\Delta_0$ for gaps according to the flow equation~(\ref{eqn:red-se}) and
 the WI~(\ref{eqn:red-WI}) for $U=1.0$.
 See text for further explanation.} \label{fig:gol-seed}
\end{figure}

 So far, we have discussed the fulfillment of the WI by looking at the values of the gap. But there is another property associated with the WI.
 In the infrared, the interaction is dominated by the Goldstone vertex and therefore other contributions on the right-hand side
 of the WI~(\ref{eqn:red-WI}) seem to be of minor importance. If the SU(2) and translational symmetries are \emph{spontaneously} broken, i.e.\ if
 the gap does not vanish for $\Delta_0 \to 0$, the Goldstone vertex must diverge as $ M_{xy} (0) \propto \Delta_0^{-1} $ in this limit.
 In Fig.~\ref{fig:gol-seed}, the reciprocal value of the Goldstone vertex is depicted for various
 values of the seed field. Simply integrating the flow equations (upper curve in Fig.~\ref{fig:gol-seed}) gives rise to points that 
 agree well with their linear fit. However, if $ 1/M_{xy} (0)$ is extrapolated to $ \Delta_0 \to 0$, one still obtains a 
 finite Goldstone vertex as a consequence of WI violations.

 The WI can now be enforced by determining $\Delta$ by iterating Eq.~(\ref{eqn:red-WI}) until convergence is reached at each iteration step
 of the ordinary differential equation (ODE) solver. [Its scale-derivative
 needed in the loops, in contrast, is still obtained from the flow equation~(\ref{eqn:red-se}).]
 The resulting infrared values of $1/M_{xy} (0)$ constitute the lower curve in Fig.~\ref{fig:gol-seed}. Again, a linear dependence on the seed field is
 found. But now, the corresponding fit curve is much closer to the origin for a vanishing seed field.
 This indicates that enforcing the WI not only somehow projects the fRG flow on the
 hypersurface in parameter  space given by this identity, but also leads to physically meaningful results.
 An even more promising approach would constitute in applying an ODE solver with a constraint \cite{coord-proj} as in Refs.~\onlinecite{eberlein-sssu,eberlein-repu}.
 Since a number of steps in a more quantitative direction still need to be undertaken before, we refrain from this task here.
 (Except for Fig.~\ref{fig:gol-seed}, the WI is \emph{not} enforced in the figures of this work, i.e., the gap is obtained from the flow equations.)
 
 Let us note in passing that $\Delta_0 $ cannot be chosen arbitrarily small before the fermionic cutoff $\lambda$ has been fully removed.
 This is due to the singular behavior of box diagrams with bosonic lines corresponding to the Goldstone vertex.
The discussion of these diagrams in Refs.~\onlinecite{eberlein-sssu,eberlein-phd} also applies for the present case and the $\Delta_0$ flow
 proposed in those works offers itself as a method for the removal of the seed field. But, before such a flow is implemented, again a considerable
 amount of work remains to be done in order to first reduce the WI violation further.
\subsection{Comparison to mean-field theory} \label{sec:comp-MF}
\begin{figure}
  \centering
  \includegraphics[width=\linewidth]{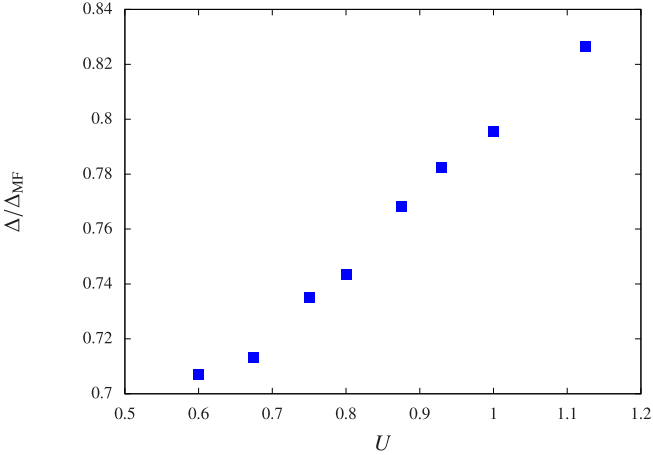}
\caption{(Color online) Ratio of the fRG and mean-field gaps $\Delta = \Delta^{\lambda=0} (0)$ and $\Delta_{\rm MF}$ as a function of $U$. See text for further explanation.
} \label{fig:mf-comp}
\end{figure}

The present analysis represents a first step beyond the mean-field picture in a fRG approach to antiferromagnetically ordered phases.
 The corresponding flow equations~(\ref{eqn:red-D})--(\ref{eqn:red-se}) reproduce the mean-field result in RPA (see Sec.~\ref{sec:rpa}).
 The fRG flow behaves RPA-like in the sense that the coupling between different channels
 induces only finite renormalizations, in analogy to the fRG flow of the attractive Hubbard model into the superfluid phase. \cite{eberlein-sssu}
  Let us have a look at these renormalizations for the two-pocket model here.
 In Fig.~\ref{fig:mf-comp}, the ratio $\Delta / \Delta_{\rm MF} $ of the gaps obtained from fRG
 and mean-field theory is plotted against the interaction strength $U$. Note that the fRG values $\Delta$  calculated
 for nonvanishing seed fields are only upper estimates for the gap. For all data points depicted, a reduction of the gap through the coupling of different channels can be observed.
 The present data suggest that $\Delta / \Delta_{\rm MF} $ increases with the interaction strength $U$. A similar increase has also been found for 
 the superconducting gap of the attractive Hubbard model in Ref.~\onlinecite{eberlein-sssu}. 

 Clearly, a reduction of the mean-field gap may also partly occur if $U^2$ terms in the gap equation~(\ref{eqn:MF-chubu}) were added.
 Such a self-consistency equation for the gap may be obtained
% from an expansion of the Luttinger-Ward functional \cite{luttinger-functional} up to second order in $U$.
 in different ways. \cite{Onsager-Hubbard,Onsager-Kopietz,AF-Schauerte}
 While methods of this kind have been used for the two-dimensional Hubbard model in Refs.~\onlinecite{Onsager-Kopietz,AF-Schauerte},
 such considerations have not yet been undertaken for the two-pocket model to the authors' knowledge.
 
 The renormalizations of the RPA result contained in the fRG values of $\Delta$ are caused by diagrams with bosonic lines \emph{inside} closed loops
 on the right-hand sides of the flow equations~(\ref{eqn:red-D})--(\ref{eqn:red-se}). The importance of these vertex-correction and box diagrams
 may manifest itself in different ways.
\begin{enumerate}[i.)]
 \item The frequency dependence of the vertex may affect the results. In static approximation, the WI could then be more strongly 
 violated.
 \item The feedback of the CDW and singlet-pairing channels on the other interaction channels and the
 gap, which is absent at the RPA level, may play a role.
 The WI should be more strongly violated if the corresponding exchange propagators $N$ and $D$ are neglected.
\end{enumerate}
\begin{figure}
  \centering
  \includegraphics[width=\linewidth]{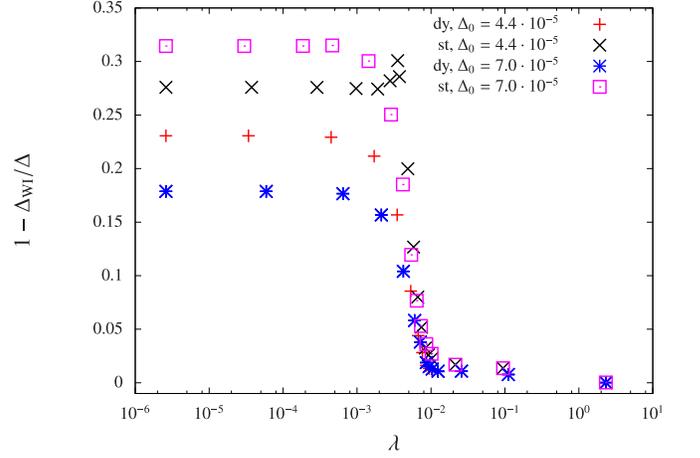}
\caption{(Color online) Violation of the WI~(\ref{eqn:red-WI}) as a function of the scale $\lambda$ for a frequency-dependent vertex (dy) and in static approximation (st).
The fRG results for the gap are denoted by $\Delta$, while $\Delta_{\rm WI}$ corresponds to the right-hand side of the WI~(\ref{eqn:red-WI}).
 The model parameters $\epsilon_0 = 3.0 \cdot 10^{-2} $ and $U =1.0$ are  kept fixed, while $\Delta_0$ is varied. 
} \label{fig:wis-nofreq}
\end{figure}

 Let us therefore first have a look at the impact of the frequency dependence, i.e. compare the flow with a frequency-dependent vertex to the 
 flow in static approximation.
 As already mentioned above, the fulfillment of the WI can be regarded as a hallmark of the quality of the approximations employed.
 In Fig.~\ref{fig:wis-nofreq}, the relative WI violation $ \left( \Delta - \Delta_{\rm MF} \right) / \Delta $ is depicted for a frequency-dependent
 vertex and in static approximation for $U=1.0$. Apparently, relaxing the frequency dependence enhances the violation of the WI. 
 As for the infrared values of the gap, the static approximation yields $ \Delta_{\rm st} = 8.52 \cdot 10^{-3} $ for $\Delta_0 = 4.4 \cdot 10^{-5} $, while 
 $ \Delta = 9.23 \cdot 10^{-3} $ is obtained with a frequency-dependent vertex.
  One may hence conclude that taking the frequency dependence of the
 vertex into account makes the present approach more powerful.
 Furthermore, the reduction of the mean-field gap is overestimated in static approximation.
 Also these findings are in agreement with those for the superfluid phase of the two-dimensional attractive Hubbard model.

\begin{figure}
  \centering
  \includegraphics[width=\linewidth]{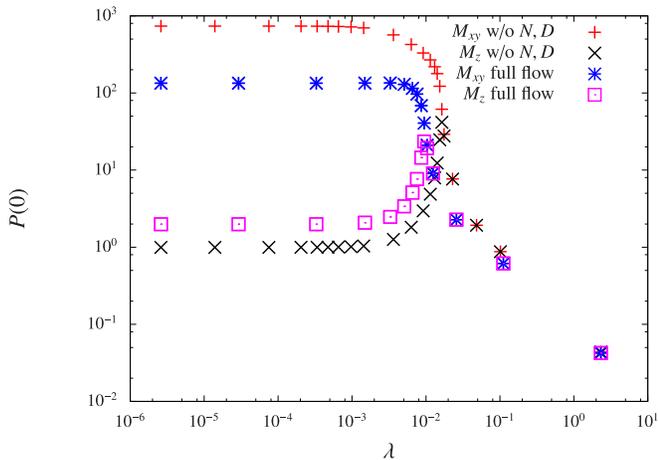}
\caption{(Color online) Comparison of the flows with and without the CDW and singlet-pairing channels for $U=1.0$ and $\Delta_0 = 5.0 \cdot 10^{-5} $.
 The exchange propagators $M_{xy} (0) $ and $M_z(0)$ are depicted as functions of the scale $\lambda$.
} \label{fig:nocdw}
\end{figure}

 Finally, let us discuss the importance of the feedback of the singlet-pairing and CDW channels on the fRG flow of the other quantities.
 The corresponding exchange propagators $D$ and $N$ are neglected for this purpose and the flow is then run for $U=1.0$ and $\Delta_0
 = 5.0 \cdot 10^{-5} $. In Fig.~\ref{fig:nocdw}, the flows with and without the singlet-pairing and CDW channels are compared.
 One can observe that without these contributions the critical scale is slightly enhanced.
 In their absence, the Goldstone vertex grows more strongly slightly above the critical scale.
 This enhanced growth of the Goldstone vertex goes along with a strong violation of the WI.
 The value $1.60 \cdot 10^{-2} $ for the gap in the absence of $D$ and $N$ considerably exceeds the fRG result $\Delta= 9.25 \cdot 10^{-3} $ in their presence,
 wrongly predicting an enhancement compared to the mean-field result $ \Delta_{\rm MF} = 1.16 \cdot 10^{-2} $.
 Altogether, this invalidates the omission of the CDW and singlet-pairing channels as a sensible approximation.
 Regarding the violation of the WI, these channels seem to be more essential than the time-reversal breaking, $s^\pm$-wave and non-momentum-conserving
 terms omitted in the
 flow equations~(\ref{eqn:red-D})--(\ref{eqn:red-se}), which \emph{a posteriori} justifies the underlying approximations.

 One may wonder why the CDW and singlet-pairing channels seem to play such an essential role. At the RPA level, the flow of these two channels does not 
 feed back on other scale-dependent quantities (see Sec.~\ref{sec:rpa}). If they are neglected, however, vertex-correction
 and box diagrams give rise to a strong enhancement of the mean-field result, which strongly violates the WI.
 Once the CDW and singlet-pairing channels with exchange propagators $N$ and $D$ are included, however, we only observe  a rather moderate reduction
 of the mean-field gap.
 This suggests that there are counteracting tendencies in the vertex-correction and box diagrams, which account for the effects beyond the mean-field picture.

 In the flow equation (\ref{eqn:red-mxy}) for $M_{xy}$,
 the linear combination $P_2=M_z/2+3D/2$ of exchange propagators appears inside the loops of these diagrams.
 Let us now recall that $D=-N$ takes on negative values, while $M_z$ is positive. Consequently, a partial cancellation of these contributions in $P_2$
 indeed reduces the impact of effects beyond the mean-field picture.
 At scales slightly above the critical scale, where $M_z (l) \simeq -3 D (l) $ for small $l$, vertex-correction and box diagrams only give negligible
 contributions to the flow of the Goldstone vertex.
 If $D$ is however neglected, the impact of these diagrams is exaggerated. This in turn gives rise to a
 strong growth of the Goldstone vertex, which results in a quite severe violation of the WI.
 Summarizing, the inclusion of the CDW and singlet-pairing channels appears to be essential for renormalizations of the mean-field result,
 while they can be neglected at the RPA level.

\section{Summary}

 In this work, we have continued fermionic  fRG flows into an antiferromagnetic phase beyond the mean-field level.
 It complements previous studies of superconductivity within a purely fermionic framework \cite{brokensy_SHML,eberlein_param,gersch_superfluid,eberlein-sssu,eberlein-repu,eberlein-phd}
 and of partially bosonized flows for various types of ordering. \cite{krahl_Hubbard_rebos,krahl_iAFM,friederich_10,friederich_11}
 In an antiferromagnet, the (discrete) translational symmetry and the (continuous) SU(2) spin symmetry are broken simultaneously.
 We have presented a physically meaningful channel decomposition of the fRG flow equations in the AF phase.
 Of course, this decomposition and the symmetry considerations made here may also be useful in other theoretical approaches where the
 vertex and the Green's functions in the symmetry-broken state constitute important building blocks.

  We have argued that,
in order to reduce the computational effort, 
 one may neglect two-particle interaction terms that break the discrete time-reversal and translational symmetries.
On the one-particle level, in contrast, these two symmetries are still broken.
 An exchange parametrization is employed and only plain $s$-wave form factors have been retained. 
 In our implementation, we have parametrized the momentum dependence of the exchange propagators by a Lorentzian in a gradient-expansion spirit, and we have discretized their frequency dependence.

 Despite all these approximations, the mean-field gap equation can be exactly reproduced from the resulting fRG flow equations in RPA,
 and, consequently, the full flow allows us to gain insight into effects beyond the mean-field picture.
 At that level, we have solved the fRG flow numerically, and our results agree with our expectation. In the present (perfectly nested) test case, only finite renormalizations compared to RPA are found.
 For these corrections to MFT, the inclusion of the CDW and singlet-pairing channel turns out to be crucial.
 The size of the gap is reduced by the contributions beyond RPA in a similar way as the superconducting gap of the attractive Hubbard model. \cite{eberlein-sssu}
 The frequency dependence of the gap is found to be negligibly weak.
 The feedback of the frequency dependence of the exchange propagators on the zero frequency couplings, in contrast, considerably improves the fulfillment of the WI. 

 This gives rise to WI violations that are small enough not to spoil our results on a qualitative level,
 which justifies our approximate parametrization \emph{a posteriori}.
 In order to obtain more precise quantitative predictions for AF gaps, future work may be geared to improving the fulfillment of the WI.
 Most likely, this can be accomplished by including the normal parts of the self-energy and the frequency dependence of the fermion-boson vertices.

 We have seen that the renormalization of the mean-field gap is predominantly due to the coupling of different interaction
 channels \emph{above} the critical scale. 
 A recently proposed fusion \cite{fRG+MF} of fRG in the symmetric phase and MFT should therefore be applicable. 
 In the superconducting phase,
 the authors of Ref.~\onlinecite{fRG+MF} have observed good agreement with symmetry-broken fRG flows. 
 For a lattice model, 
 a quantitative comparison should also be carried out for the antiferromagnetic case, and the outcome of this work suggests positive results.
 In the long run, the method used here should be applied also to the Hubbard model and other model Hamiltonians of interest.
 In addition, our channel decomposition in the general form presented in Sec.~\ref{sec:AFM-ch-dec} may also be useful for instability analyses of
 models that have a collinear $\mathrm{U}_z (1)$ instead of a full $\mathrm{SU} (2)$ spin symmetry.\footnote{%
 Recently, models without the full $\mathrm{SU} (2) $ spin symmetry have increasingly attracted research interest.
For example, a fRG study has been carried out for the  Kitaev-Heisenberg model on the honeycomb lattice, \cite{fRG-Kitaev} which, however, does not even have a $\mathrm{U}_z (1)$ spin symmetry.}
 By dropping the Nambu indices, it can also be applied to problems without breaking of the translational symmetry, e.g.\ with just a spin-splitting term. 
 In this context,
 the Kane-Mele-Hubbard model \cite{QSH-QMC,KMH-signprob,QPT-KMH} might be an interesting candidate.

\begin{acknowledgments}
 We thank K.-U.~Giering, T.~Holder, B.~Obert, M.~Salmhofer, and M.~M.~Scherer for valuable discussions. This project was supported by the German Research Foundation DFG via
 FOR 723.
\end{acknowledgments}

\begin{appendix}
 \section{Channel decomposition} \label{sec:ch-dec}
In the fRG flow equations, a direct and unbiased discretization of the coupling functions $ V_{\uparrow \downarrow} $, $V_\uparrow$ and $V_\downarrow$
 defined in Sec.~\ref{sec:gen-context} would either require further approximations, such as projection to the Fermi surface and to zero frequency, or result in even more prohibitive numerical effort than in the $ \mathrm{SU} (2) $ symmetric case. 
Therefore a so-called channel decomposition of the interaction as pioneered in Refs.~\onlinecite{karrasch,husemann_09} seems appropriate.
 Recently, such a decomposition was proposed \cite{eberlein_param} and implemented \cite{eberlein-sssu} for singlet superconductors.
 We now present such
a channel decomposition for AF phases, where not only the $ \mathrm{SU} (2) $ symmetry, but also the translational symmetry is broken.
\subsection{Formal decomposition} \label{sec:form-deco}
As we have already discussed in Ref.~\onlinecite{me-AFM-MF},
the three coupling functions from Sec.~\ref{sec:param-gen} can now be decomposed as follows.
Renormalizations of  equal-spin interactions $ W_\uparrow $ and $ W_\downarrow $ can be regarded as a sum $ \Phi_{\mathrm{SC} \sigma} $
 of triplet and anomalous pairing terms and a spin-dependent particle-hole  term $ \Phi_{K \sigma} $, which enter according to
\begin{align*}
 W_\sigma (K_1,K_2,K_3,K_4)  = & \, \tilde{\delta}_{\left\{ k_i \right\} } \left[ U_\sigma^{\left\{ s \right\} }  (k_1,k_2,k_3) \right.\\
 & + \Phi_{\mathrm{SC},\sigma}^{\left\{ s \right\} }  (k_1+k_2,k_1,k_3) \\ \notag & - \Phi_{K,\sigma}^{\left\{ s \right\} }  (k_1-k_3,k_1,k_2)\\ & + \left. \Phi_{K,\sigma}^{\left\{ \tilde{s} \right\} }  (k_3-k_2,k_1,k_2) \right] \, .
\end{align*}
In this equation, $ U_\sigma $ stems from the bare interaction and $\{ \tilde{s}\} $ denotes $(s_1,s_2,s_4,s_3) $.
The particle-hole part $ \Phi_{K,\sigma} $ contains $ S_z^2 $ and $ n^2 $ (CDW) contributions as well as terms of $S_z n$ form, 
where $n$ represents the charge density.
 In contrast to $W_\sigma$, the single-channel coupling functions $ \Phi_{\dots}$ depend strongly on one momentum and frequency argument and weakly on the other
 two. This way, the discretization effort is reduced from $ N^3$ to $N$.

 The coupling function $ W_{\uparrow \downarrow} $ with bare values $ U_{\uparrow \downarrow} $ is renormalized by a particle-particle part $\Phi_{\mathrm{SC}, \uparrow \downarrow}$, which may contain triplet, singlet and anomalous 
 pairing terms, and magnetic contributions $ \Phi_\mathrm{plane} $ corresponding to $ S_x^2 + S_y^2 $ or $ S_x S_y $ and $\Phi_\mathrm{axis} $, which contains $S_z^2$, CDW and $ S_z n$ terms 
\begin{align*}
 W_{\uparrow \downarrow} (K_1,K_2,K_3,K_4)  =  & \, \tilde{\delta}_{\left\{ k_i \right\} } \left[ U_{\uparrow \downarrow}^{\left\{ s \right\} }  (k_1,k_2,k_3) 
 \right.\\ &  + \Phi_{\mathrm{SC}, \uparrow \downarrow}^{\left\{ s \right\} }  (k_1+k_2,k_1,k_3) \\
&+  \Phi_\mathrm{plane}^{\left\{ s \right\} }  (k_3-k_2,k_1,k_2)\\ & \left.- \Phi_\mathrm{axis}^{\left\{ s \right\} }  (k_1-k_3,k_1,k_2)\right] \, . 
\end{align*}
 In Ref.~\onlinecite{me-AFM-MF}, we have derived fRG flow equations for the single-channel coupling functions $\Phi_{\dots}$ and the self-energy $\Sigma$, which we recapitulate in the following. A dot then denotes the derivate with respect to the infrared cutoff $\lambda$. Note that the following flow equations hold irrespective
 of the precise form of the regularization scheme.

In the particle-particle channels, one obtains
\begin{widetext}
\begin{align} \label{eqn:flow-Phi-tSC}
 \dot{\Phi}_{\mathrm{SC}, \sigma}^{\left\{ s \right\}} (l,q,q') & = \frac{1}{2} \sum_{\left\{ s'_i \right\}} \int \!\! d' p \, W_\sigma^{s_1,s_2,s'_1,s'_3} (q,l-q,p_+,-p_-) \,
 W_\sigma^{s'_4,s'_2,s_3,s_4} (-p_-,p_+,q',l-q') \, L_{\sigma,\sigma}^{\left\{ s'_i \right\} } (p_+,-p_-)\, , \\ \label{eqn:flow-Phi-SC} 
 \dot{\Phi}_{\mathrm{SC}, \uparrow \downarrow}^{\left\{ s \right\}} (l,q,q') & = - \sum_{\left\{ s'_i \right\}} \int \!\! d' p \, W_{\uparrow \downarrow}^{s_1,s_2,s'_1,s'_3} (q,l-q,p_+,-p_-) \, W_{\uparrow \downarrow}^{s'_2,s'_4,s_3,s_4} (p_+,-p_-,q',l-q') \, L_{\uparrow,\downarrow}^{\left\{ s'_i \right\} } (p_+,-p_-)  \, ,
\end{align}
where $p_\pm = p \pm l/2 $.
The flow in the particle-hole channels is governed by
 \begin{align} \notag
  \dot{\Phi}_{K, \uparrow}^{\left\{ s \right\}} (l,q,q') & = -\sum_{\left\{ s'_i \right\}} \int \!\! d' p \, W_\uparrow^{s'_4,s_2,s_4,s'_1} (p_+,q',l+q',p_-) 
 \, W_\uparrow^{s_1,s'_2,s'_3,s_3} (q,p_-,p_+,q-l) \, L_{\uparrow,\uparrow}^{\left\{ s'_i \right\} } (p_-,p_+)
 \\  \label{eqn:flow-Phi-Ku} 
& \quad -\sum_{\left\{ s'_i \right\}} \int \!\! d' p \, 
 W_{\uparrow \downarrow}^{s_2,s'_4,s_4,s'_1} (q',p_+,l+q',p_-) \, W_{\uparrow \downarrow}^{s_1,s'_2,s_3,s'_3} (q,p_-,q-l,p_+)  \, L_{\downarrow,\downarrow}^{\left\{ s'_i \right\} } (p_-,p_+)  \, ,
 \end{align}
 \begin{align} \notag
  \dot{\Phi}_{K, \downarrow}^{\left\{ s \right\}} (l,q,q') & = - \sum_{\left\{ s'_i \right\}} \int \!\! d' p \, W_\downarrow^{s'_4,s_2,s_4,s'_1} (p_+,q',l+q',p_-) \,
W_\downarrow^{s_1,s'_2,s'_3,s_3} (q,p_-,p_+,q-l) \, L_{\downarrow,\downarrow}^{\left\{ s'_i \right\} } (p_-,p_+)  \\ \label{eqn:flow-Phi-Kd} 
& \quad - \sum_{\left\{ s'_i \right\}} \int \!\! d' p \, W_{\uparrow \downarrow}^{s'_4,s_2,s'_1,s_4} (p_+,q',p_-,l+q') \, W_{\uparrow \downarrow}^{s'_2,s_1,s'_3,s_3} (p_-,q,p_+,q-l)  \, L_{\uparrow,\uparrow}^{\left\{ s'_i \right\} } (p_-,p_+)  \, ,
 \end{align}
\begin{equation} \label{eqn:flow-Phi-plane} 
  \dot{\Phi}_\mathrm{plane}^{\left\{ s \right\}} (l,q,q') = -  \sum_{\left\{ s'_i \right\}} \int \!\! d' p \, W_{\uparrow \downarrow}^{s'_4,s_2,s_3,s'_1} (p_+,q',l+q',p_-) \, W_{\uparrow \downarrow}^{s_1,s'_2,s'_3,s_4} (q,p_-,p_+,q-l) \, L_{\downarrow,\uparrow}^{\left\{ s'_i \right\} } (p_-,p_+)  \, ,
\end{equation}
 \begin{align*}
  \dot{\Phi}_\mathrm{axis}^{\left\{ s \right\}} (l,q,q') & = \sum_{\left\{ s'_i \right\}} \int \!\! d' p \, W_{\uparrow \downarrow}^{s'_4,s_2,s'_1,s_4} (p_+,q',p_-,q'+l) \, W_\uparrow^{s_1,s'_2,s'_3,s_3} (q,p_-,p_+,q-l) \, L_{\uparrow,\uparrow}^{\left\{ s'_i \right\} } (p_-,p_+) \\
 & \quad +\sum_{\left\{ s'_i \right\}} \int \!\! d' p \, W_\downarrow^{s'_4,s_2,s_4,s'_1} (p_+,q',l+q',p_-) \, W_{\uparrow \downarrow}^{s_1,s'_2,s_3,s'_3} (q,p_-,q-l,p_+) \, L_{\downarrow,\downarrow}^{\left\{ s'_i \right\} } (p_-,p_+)  \, .
 \end{align*}
 Expressed in Nambu space, the flow  equations for the self-energy read as
\begin{align*}
 \partial_\lambda \Sigma^{s_1 s_2}_\uparrow (k_1,k_2)  & = - \sum_{s'_1 s'_2} \int \! d' p  \, S^{s'_1 s'_2}_\uparrow (p) \, W^{s_1,s'_2,s'_1,s_2}_\uparrow (k_1,p,p,k_2)
+ \sum_{s'_1 s'_2} \int \! d' p  \,  S^{s'_1 s'_2}_\downarrow (p) \, W^{s_1,s'_2,s_2,s'_1}_{\uparrow\downarrow} (k_1,p,k_2,p) \\
 \partial_\lambda \Sigma^{s_1 s_2}_\downarrow (k_1,k_2)  &= - \sum_{s'_1 s'_2} \int \! d' p \, S^{s'_1 s'_2}_\downarrow (p) \, W^{s_1,s'_2,s'_1,s_2}_\downarrow (k_1,p,p,k_2) 
 + \sum_{s'_1 s'_2} \int \! d' p  \,  S^{s'_1 s'_2}_\uparrow (p) \, W^{s'_2,s_1,s'_1,s_2}_{\uparrow\downarrow} (p,k_1,p,k_2) \, ,
\end{align*}
\end{widetext}
 where the single-scale propagator defined in Eq.~(\ref{eqn:def-sscale}) is equal to the scale derivative
 \begin{equation*}
  S^{s s'}_\sigma (k) = \partial_\lambda \left. G^{s s'}_\sigma (k) \right|_{\Sigma= \mathrm{const}}
 \end{equation*}
 of the one-particle propagator with the self-energy held constant.

 \subsection{Improved parametrization}  \label{sec:improved-param}

 In the present form, this channel decomposition would already allow for a reduction of computational effort if all three momentum and frequency
 variables were discretized. This would, however, rather constitute an approximation simplifying the numerics than a decomposition into physically meaningful channels. 
 Namely, $ \Phi_K $ and $\Phi_\mathrm{axis} $ both contain $S_z^2$ and CDW contributions.
 In a physically meaningful channel decomposition that allows for sensible further approximations, however,
 $S_z^2$ and CDW contributions should appear in different channels. 
 In the following, this will be accomplished by decomposing the single-channel coupling functions into spin-normal and spin-anomalous contributions
 and then linearly recombining the spin-normal parts.
 
 Let us first decompose $ \Phi_\mathrm{axis} $ into its spin-normal and spin-anomalous parts
\begin{equation*}
 \Phi^{\{s\}}_{\mathrm{axis} \pm} (l,p,q) = 
 \frac{1}{2} \left[\Phi^{\{s\}}_\mathrm{axis} (l,p,q) \pm \Phi^{\{\bar{s}\}}_\mathrm{axis} (-l,q,p) \right] \, , 
\end{equation*}
 where $ \{ \bar{s} \} = (s_2,s_1,s_4,s_3) $.
 With the short-hand notation
\begin{equation*}
 L_{\sigma_1,\sigma_2}^{\left\{ s \right\} } (p,q) = \partial_\lambda \left[ G_{\sigma_1}^{s_1,s_2} (p) \,G_{\sigma_2}^{s_3,s_4} (q) \right]
\end{equation*}
 for the loops, their scale derivatives can be cast into the form
\begin{widetext}
 \begin{align} \notag
  \dot{\Phi}_{\mathrm{axis} \pm}^{\left\{ s \right\}} (l,q,q') & = \frac{1}{2}  \sum_{\left\{ s'_i \right\}} \int \!\! d' p \, W_{\uparrow \downarrow}^{s'_4,s_2,s'_1,s_4} (p_+,q',p_-,q'+l) \, W_\uparrow^{s_1,s'_2,s'_3,s_3} (q,p_-,p_+,q-l) \, L_{\uparrow,\uparrow}^{\left\{ s'_i \right\} } (p_-,p_+) \\ \notag
 & \quad \pm \frac{1}{2} \sum_{\left\{ s'_i \right\}} \int \!\! d' p \, W_{\uparrow \downarrow}^{s_2,s'_4,s_4,s'_1} (q',p_+,q'+l,p_-) \, W_\downarrow^{s_1,s'_2,s'_3,s_3} (q,p_-,p_+,q-l) \, L_{\downarrow,\downarrow}^{\left\{ s'_i \right\} } (p_-,p_+) \\ \notag
 & \quad +  \frac{1}{2}  \sum_{\left\{ s'_i \right\}} \int \!\! d' p \, W_\downarrow^{s'_4,s_2,s_4,s'_1} (p_+,q',q'+l,p_-) \, W_{\uparrow \downarrow}^{s_1,s'_2,s_3,s'_3} (q,p_-,q-l,p_+) \, L_{\downarrow,\downarrow}^{\left\{ s'_i \right\} } (p_-,p_+)  \\ \label{eqn:flow-Phi-ax-}
  & \quad \pm \frac{1}{2} \sum_{\left\{ s'_i \right\}} \int \!\! d' p \, W_\uparrow^{s'_4,s_2,s_4,s'_1} (p_+,q',q'+l,p_-) \, W_{\uparrow \downarrow}^{s'_2,s_1,s'_3,s_3} (p_-,q,p_+,q-l) \, L_{\uparrow,\uparrow}^{\left\{ s'_i \right\} } (p_-,p_+) \, . 
 \end{align}
 Likewise, one may introduce spin-normal and spin-anomalous coupling functions
\begin{align*}
%\Phi^{\{s\}}_{\mathrm{SC} \pm} (l,p,q) & = 
%\frac{1}{2} \left[\Phi^{\{s\}}_{\mathrm{SC}, \uparrow } (l,p,q) \pm \Phi^{\{\bar{s}\}}_{\mathrm{SC},\downarrow } (l,p,q) \right] \, , \\
 \Phi^{\{s\}}_{K \pm} (l,p,q) & = \frac{1}{2} \left[\Phi^{\{s\}}_{K, \uparrow } (l,p,q) \pm \Phi^{\{\bar{s}\}}_{K,\downarrow } (l,p,q) \right] 
\end{align*}
 for the %tSC and
 $K$ channels.
 Their scale derivative can be obtained by adding or subtracting the flow equations~(\ref{eqn:flow-Phi-Ku}) and (\ref{eqn:flow-Phi-Kd}), respectively.
The corresponding flow equations now read as follows:
 \begin{align} \notag
  \dot{\Phi}_{K, \pm}^{\left\{ s \right\}} (l,q,q') & = - \frac{1}{2}  \sum_{\left\{ s'_i \right\}} \int \!\! d' p \, W_\uparrow^{s'_4,s_2,s_4,s'_1} (p_+,q',l+q',p_-) \, W_\uparrow^{s_1,s'_2,s'_3,s_3} (q,p_-,p_+,q-l) \, L_{\uparrow,\uparrow}^{\left\{ s'_i \right\} } (p_-,p_+)  \\ \notag 
  & \quad \mp  \frac{1}{2}  \sum_{\left\{ s'_i \right\}} \int \!\! d' p \, W_\downarrow^{s'_4,s_2,s_4,s'_1} (p_+,q',l+q',p_-) \, W_\downarrow^{s_1,s'_2,s'_3,s_3} (q,p_-,p_+,q-l) \, L_{\downarrow,\downarrow}^{\left\{ s'_i \right\} } (p_-,p_+)  \\ \notag 
& \quad  -  \frac{1}{2}  \sum_{\left\{ s'_i \right\}} \int \!\! d' p \, W_{\uparrow \downarrow}^{s_2,s'_4,s_4,s'_1} (q',p_+,l+q',p_-) \, W_{\uparrow \downarrow}^{s_1,s'_2,s_3,s'_3} (q,p_-,q-l,p_+)  \, L_{\downarrow,\downarrow}^{\left\{ s'_i \right\} } (p_-,p_+)  \\ \label{eqn:flow-Phi-K-} 
& \quad  \mp  \frac{1}{2}  \sum_{\left\{ s'_i \right\}} \int \!\! d' p \, W_{\uparrow \downarrow}^{s'_4,s_2,s'_1,s_4} (p_+,q',p_-,l+q') \, W_{\uparrow \downarrow}^{s'_2,s_1,s'_3,s_3} (p_-,q,p_+,q-l)  \, L_{\uparrow,\uparrow}^{\left\{ s'_i \right\} } (p_-,p_+)  \, ,
 \end{align}
 $ S_z^2 $ and CDW contributions are spin normal and can be obtained as 
\begin{equation*}
 \Phi_z^{\{s\}} (l,p,q) = \Phi_{K +}^{\{s\}} (l,p,q) - \Phi_{\mathrm{axis} +}^{\{s\}} (l,p,q) \, , \qquad 
 \Phi_\mathrm{CDW}^{\{s\}} (l,p,q) = \Phi_{K +}^{\{s\}} (l,p,q) + \Phi_{\mathrm{axis} +}^{\{s\}} (l,p,q) \, ,
\end{equation*}
 respectively, as illustrated in Fig.~\ref{fig:new-props}. With the shorthand notations
\begin{equation} \label{eqn:pm-shorthands}
 W_{\pm \sigma}^{\{s\}} (k_1,k_2,k_3,k_4) = W_{\uparrow \downarrow}^{\{s\}} (k_1,k_2,k_3,k_4) \pm W_\sigma^{\{\tilde{s}\}} (k_1,k_2,k_4,k_3)  \, ,
\end{equation}
the flow equations of these new single-channel coupling functions read as follows.
 Once again, the prime in the measure $ d'p $ indicates that the respective momentum integral only runs over the magnetic BZ:
 \begin{align} \notag
 \dot{\Phi}_\mathrm{CDW}^{\{s\}} (l,q,q') = & \, - \frac{1}{2}  \sum_{\left\{ s'_i \right\}} \int \!\! d' p \, L_{\uparrow,\uparrow}^{\left\{ s'_i \right\} } (p_-,p_+) \,
W_{- \uparrow}^{s'_2,s_1,s'_3,s_3} (p_-,q,p_+,q-l) \,
   W_{- \uparrow}^{s'_4,s_2,s'_1,s_4} (p_+,q',p_-,q'+l) \\  \label{eqn:flow-Phi-CDW}
 &- \frac{1}{2}  \sum_{\left\{ s'_i \right\}} \int \!\! d' p \, L_{\downarrow,\downarrow}^{\left\{ s'_i \right\} } (p_-,p_+) 
 \, W_{- \downarrow}^{s_1,s_2',s_3,s'_3} (q,p_-,q-l,p_+) \,
   W_{- \downarrow}^{s_2,s'_4,s_4,s'_1} (q',p_+,q'+l,p_-) 
 \end{align}
 and
 \begin{align} \notag
 \dot{\Phi}_z^{\{s\}} (l,q,q') = & \, - \frac{1}{2}  \sum_{\left\{ s'_i \right\}} \int \!\! d' p \, L_{\uparrow,\uparrow}^{\left\{ s'_i \right\} } (p_-,p_+) \, W_{+ \uparrow}^{s'_2,s_1,s'_3,s_3} (p_-,q,p_+,q-l) \,
   W_{+ \uparrow}^{s'_4,s_2,s'_1,s_4} (p_+,q',p_-,q'+l) \\ \label{eqn:flow-Phi-z}
 &- \frac{1}{2}  \sum_{\left\{ s'_i \right\}} \int \!\! d' p \, L_{\downarrow,\downarrow}^{\left\{ s'_i \right\} } (p_-,p_+) \, W_{+ \downarrow}^{s_1,s_2',s_3,s'_3} (q,p_-,q-l,p_+) \,
   W_{+ \downarrow}^{s_2,s'_4,s_4,s'_1} (q',p_+,q'+l,p_-)  \, .
 \end{align}
 In the more physical parametrization presented here, the single-channel coupling functions $ W_\uparrow $, $ W_\downarrow $, and $ W_{\uparrow \downarrow} $
 are decomposed as follows:
\begin{align*} 
 W_\uparrow (K_1,K_2,K_3,K_4)  = & \, \tilde{\delta}_{\left\{ k_i \right\} } \left[ U_\uparrow^{\left\{ s \right\} }  (k_1,k_2,k_3) + \Phi_{\mathrm{SC},\uparrow}^{\left\{ s \right\} }  (k_1+k_2,k_1,k_3)
- \frac{1}{2}\Phi_\mathrm{CDW}^{\left\{ s \right\} }  (k_1-k_3,k_1,k_2) \right. \\ & 
- \frac{1}{2}\Phi_{z}^{\left\{ s \right\} }  (k_1-k_3,k_1,k_2)
 - \Phi_{K-}^{\left\{ s \right\} }  (k_1-k_3,k_1,k_2)
 + \frac{1}{2}\Phi_\mathrm{CDW}^{\left\{ \tilde{s} \right\} }  (k_3-k_2,k_1,k_2) \\
 &+ \frac{1}{2}\Phi_{z}^{\left\{ \tilde{s} \right\} }  (k_3-k_2,k_1,k_2) 
  + \left.  \Phi_{K-}^{\left\{ \tilde{s} \right\} }  (k_3-k_2,k_1,k_2) \right] \, ,
\end{align*}
\begin{align*} 
 W_\downarrow (K_1,K_2,K_3,K_4)  = & \, \tilde{\delta}_{\left\{ k_i \right\} } \left[ U_\downarrow^{\left\{ s \right\} }  (k_1,k_2,k_3) + \Phi_{\mathrm{SC},\downarrow}^{\left\{ s \right\} }  (k_1+k_2,k_1,k_3)
- \frac{1}{2}\Phi_\mathrm{CDW}^{\left\{ s \right\} }  (k_1-k_3,k_1,k_2) \right. \\ & 
- \frac{1}{2}\Phi_{z}^{\left\{ s \right\} }  (k_1-k_3,k_1,k_2)
 + \Phi_{K-}^{\left\{ s \right\} }  (k_1-k_3,k_1,k_2)
 + \frac{1}{2}\Phi_\mathrm{CDW}^{\left\{ \tilde{s} \right\} }  (k_3-k_2,k_1,k_2) \\
 &+ \frac{1}{2}\Phi_{z}^{\left\{ \tilde{s} \right\} }  (k_3-k_2,k_1,k_2) 
  - \left.  \Phi_{K-}^{\left\{ \tilde{s} \right\} }  (k_3-k_2,k_1,k_2) \right] \, ,
\end{align*}
 where $ \Phi_{K-} $ enters with different signs in $ W_\uparrow $, and $ W_\downarrow$, and
\begin{align*}
 W_{\uparrow \downarrow} (K_1,K_2,K_3,K_4)  =  & \, \tilde{\delta}_{\left\{ k_i \right\} } \left[ U_{\uparrow \downarrow}^{\left\{ s \right\} }  (k_1,k_2,k_3) + \Phi_{\mathrm{SC}, \uparrow \downarrow}^{\left\{ s \right\} }  (k_1+k_2,k_1,k_3)
+ \Phi_\mathrm{plane}^{\left\{ s \right\} }  (k_3-k_2,k_1,k_2) \right. \\ &
 \left. - \frac{1}{2} \Phi_\mathrm{CDW}^{\left\{ s \right\} }  (k_1-k_3,k_1,k_2)
+  \frac{1}{2} \Phi_z^{\left\{ s \right\} }  (k_1-k_3,k_1,k_2)
 - \Phi_{\mathrm{axis}-}^{\left\{ s \right\} }  (k_1-k_3,k_1,k_2)\right] \, . 
\end{align*}
\end{widetext}
 The scale dependence of $ \Phi_\mathrm{CDW} $, $\Phi_z $, $ \Phi_{K-} $, and $ \Phi_{\mathrm{axis}-} $ is governed by the flow equations~(\ref{eqn:flow-Phi-CDW}),
 (\ref{eqn:flow-Phi-z}), (\ref{eqn:flow-Phi-K-}), and (\ref{eqn:flow-Phi-ax-}).
 In contrast, the single-channel coupling functions $ \Phi_{\mathrm{SC},\sigma} $, $ \Phi_{\mathrm{SC}, \uparrow \downarrow} $, and $ \Phi_\mathrm{plane} $ still flow according to Eqs.~(\ref{eqn:flow-Phi-tSC}), 
 (\ref{eqn:flow-Phi-SC}), and (\ref{eqn:flow-Phi-plane}).
 In a way similar to the above extraction of CDW and $S_z^2$ contributions, $ \Phi_\mathrm{plane} $ can be decomposed into
 (spin-normal) $S_x^2+S_y^2$ and (spin-anomalous) $S_x S_y$ terms. Also, singlet, triplet-, and anomalous pairing terms could be extracted from
 $ \Phi_{\mathrm{SC},\uparrow} $, $ \Phi_{\mathrm{SC},\downarrow} $, and $ \Phi_{\mathrm{SC}, \uparrow \downarrow} $.
 For the symmetries of the perfectly nested case, however, only the more important singlet-pairing contributions will be retained in Appendix~\ref{sec:time-normal-app}.

In summary, the channel decomposition presented here paves the road to an efficient (approximate) parametrization of the interaction 
resulting in numerically tractable flow equations as in Sec.~\ref{sec:AFM-fleq}, where an exchange parametrization will be employed. 
 The group-theoretic view on exchange parametrizations presented in Ref.~\onlinecite{me-Emery} also applies to the above
 flow equations for collinear spin ordering.

\section{Derivation of the fRG flow equations in exchange parametrization} \label{sec:deriv-ex-para}

 \subsection{Symmetries in the presence of perfect nesting}  \label{sec:AFM-symm}
 In an attempt to go beyond a mean-field approach, one
 is left with the full channel-decomposed flow equations of Appendix~\ref{sec:improved-param}.
 Due to the Nambu-index dependence of the coupling functions, a direct discretization of their arguments
 would still be far too costly. Therefore additional symmetries, such as the one stemming from a perfectly nested
 dispersion should be exploited in the parametrization of the coupling functions.
In this section, we will therefore discuss these symmetries for a general action of the form (\ref{eqn:action})  before incorporating them into the parametrization.
 A bare action of the type of Eq.~(\ref{eqn:action}) usually corresponds to a Hermitian Hamiltonian. This translates
 to the constraints
\begin{align}  \label{eqn:OSP-const-quad}
 C_\sigma (k,k') &= C_\sigma (\hat{k}',\hat{k})^\ast \, , \\  \label{eqn:OSP-const-quar}
 f (\xi_1,\xi_2,\xi_3,\xi_4) &= f (\hat{\xi}_4,\hat{\xi}_3,\hat{\xi}_2,\hat{\xi}_1)^\ast
\end{align}
 on the coupling functions of the action $ \mathcal{A} $ in Eq.~(\ref{eqn:action}),
 where $ \hat{k} = (-k_0,\mathbf{k}) $ and $\hat{\xi} = (\hat{k},\sigma)$.
 Note that this \emph{Osterwalder-Schrader positivity} \cite{osterwalder-schrader,wetterich-positivity} (OSP) is referred to as particle-hole symmetry in Refs.~\onlinecite{husemann_09,HGS-freq-dep,giering-new}, which
 should not be confused with the particle-hole symmetry of the Hubbard model at perfect nesting as defined in Ref.~\onlinecite{SO4-hubbard}.

 In the positivity constraint~(\ref{eqn:OSP-const-quad}) for the quadratic part of action, the \emph{Pauli principle} is already included.
 The one for the quartic part, however, is complemented by the Pauli-principle constraint
\begin{equation*}
 f (\xi_1,\xi_2,\xi_3,\xi_4) = - f (\xi_2,\xi_1,\xi_3,\xi_4) = - f (\xi_1,\xi_2,\xi_4,\xi_3) \, .
\end{equation*}

 In a fRG framework, symmetries can only be exploited if they can be formulated in terms of symmetry constraints on the coupling functions,
 Eqs.~(\ref{eqn:OSP-const-quad}) and (\ref{eqn:OSP-const-quar}) being examples thereof.
 It therefore seems worthwhile to look for such a constraint  on the coupling functions that arises from perfect nesting.
 This constraint could then be used in a further parametrization of the coupling functions.
 As for the Hubbard model with hopping only between nearest neighbors, 
 we find for the two-pocket model of Sec.~\ref{sec:chubu} that
\begin{align}\label{eqn:PHS-symmetry-quad}
 C_\sigma (k,k') & = - C^\ast_{- \sigma} (k+Q,k'+Q) \, ,\\ \label{eqn:PHS-symmetry-int}
 f (\xi_1,\xi_2,\xi_3,\xi_4) & = f^\ast (\tilde{\xi}_1,\tilde{\xi}_2,\tilde{\xi}_3,\tilde{\xi}_4) \, ,
\end{align}
 where $ \tilde{\xi}_i = \left( - \sigma_i, k_i + Q \right) $.
 In the language of Refs.~\onlinecite{SO4-hubbard,SO6-Hubb}, this corresponds to flipping the components of the pseudospinors
\begin{equation} \label{eqn:pseudospin}
 \Psi_\mathrm{p}^s (k) = \left( \begin{array}{c} \psi_\uparrow^s (k) \\ \bar{\psi}_\downarrow^{-s} (-k) \end{array} \right) \, .
\end{equation}
 This symmetry constitutes a subgroup of the `hidden' SU(2) pseudospin symmetry. On the one particle level, pseudospin-flip invariance already implies
 a pseudospin SU(2) symmetry.
 For two-particle and higher-order interaction terms, this is no longer the case.
 Since a general pseudospin rotation mixes ingoing and outgoing
 fields, fully exploiting this hidden symmetry in the parametrization of the interaction represents a challenging task, which we leave for future work.

 The symmetry constraints (\ref{eqn:PHS-symmetry-quad}) and (\ref{eqn:PHS-symmetry-int}) still hold in the presence of a nonvanishing antiferromagnetic seed field $\Delta$.
 
 But, as soon as the two-pocket model of Sec.~\ref{sec:chubu} was doped away from perfect nesting, Eq.~(\ref{eqn:PHS-symmetry-quad}) would be violated as well.
Once they are met by the bare action, the constraints in Eqs.~(\ref{eqn:PHS-symmetry-quad}) and (\ref{eqn:PHS-symmetry-int})
 will, however, be preserved by the fRG flow equations.
For the spin-independent coupling functions, the second of these constraints translates to
\begin{align*}
 V_\uparrow (k_1,k_2,k_3,k_4) & = V^\ast_\downarrow (k_1+Q,k_2+Q,k_3+Q,k_4+Q) \, ,\\
 V_{\uparrow \downarrow} (k_1,k_2,k_3,k_4) & = V^\ast_{\uparrow \downarrow} (k_2+Q,k_1+Q,k_4+Q,k_3+Q) \, ,
\end{align*}
while the first one implies a form
\begin{equation} \label{eqn:Cform}
 \begin{aligned}
 \mathbf{C}_\uparrow (k) &= \Delta (k) \, \tau^1 + \left[ i k_0 - \epsilon_\mathrm{s} (k) \right] \tau^0 - \epsilon_\mathrm{a} (k) \, \tau^3\, ,\\
 \mathbf{C}_\downarrow (k) &= - \Delta^\ast (k) \, \tau^1 + \left[ i k_0 + \epsilon_\mathrm{s}^\ast (k) \right] \tau^0 - \epsilon_\mathrm{a}^\ast (k) \, \tau^3
 \end{aligned}
\end{equation}
of the quadratic part of the action in Nambu space with spinors according to Eq.~(\ref{eqn:Nambu-spinor}).
 In the following, we will refer to this symmetry as a particle-hole symmetry (PHS).

 In Eq.~(\ref{eqn:Cform}), the normal self-energy enters with its Nambu-index symmetric and antisymmetric parts in $ \epsilon_\mathrm{s}$ and $\epsilon_\mathrm{a}$,
 respectively.
 For a bare action with a perfectly nested dispersion, $ \epsilon_\mathrm{s} = 0 $ and hence the Nambu-symmetric part of the self-energy is created, if it
 is nonvanishing at all, during the flow.
 In Nambu space, this corresponds to a propagator of the form
\begin{equation} \label{eqn:propsymm} 
 \begin{aligned}
 \mathbf{G}_\uparrow (k) & = \frac{1}{k_0^2 + 2i k_0 \epsilon_\mathrm{s} (k) - \epsilon_\mathrm{s} (k)^2 + \epsilon_\mathrm{a} (k)^2 
+ \Delta (k)^2} \\ & \, \, \times
\left( \begin{array}{cc}
 -i k_0 + \epsilon_\mathrm{s} (k) - \epsilon_\mathrm{a} (k) & -\Delta (k) \\
 -\Delta (k) & -i k_0 + \epsilon_\mathrm{s} (k) + \epsilon_\mathrm{a} (k) 
\end{array} \right)
 \end{aligned}
\end{equation}
for spin up and likewise for spin down with the substitutions $ \Delta (k) \to -\Delta^\ast (k) $,
 $ \epsilon_\mathrm{s} (k) \to - \epsilon_\mathrm{s}^\ast (k) $ and
 $ \epsilon_\mathrm{a} (k) \to \epsilon_\mathrm{a}^\ast (k) $.

 This general pseudospin-flip-symmetric form of the propagator differs from the one in the mean-field case. 
 For one thing, the bare dispersion in $ \epsilon_\mathrm{a} $ gets renormalized by contributions of the normal self-energy, which
depend on momentum and frequency. Also the anomalous part $ \Delta $ of the self-energy may show such a dependence.
 Furthermore, contributions $ \epsilon_\mathrm{s} $ to the normal self-energy appear that are symmetric under
 a Nambu index flip.
 As can be seen from the denominator of Eq.~(\ref{eqn:propsymm}), a nonvanishing value of
$ \epsilon_\mathrm{s} $ might give rise to a Fermi surface reconstruction,
 since it may cause zeros of the denominator in presence of an antiferromagnetic gap.
 Keeping track of this effect may, however, require a good momentum resolution of the self-energy within an unbiased 
 discretization scheme. In this work, we will therefore have to refrain from such tasks.
 
 In the following, other symmetries will turn out to be useful.
 Let us recall that we consider a bare action equivalent to a model Hamiltonian.
 Under frequency-inversion $ k_0 \to - k_0$, the coupling functions both in the quadratic and the quartic parts
 of the action are then mapped to their complex conjugates, i.e.,
\begin{align*}
 C_\sigma (\hat{k},\hat{k}') &= C^\ast_\sigma ({k},{k}') \, ,\\
 V_\uparrow (\hat{k}_1,\hat{k}_2,\hat{k}_3,\hat{k}_4) &= V^\ast_\uparrow ({k}_1,{k}_2,{k}_3,{k}_4) \, ,\\
 V_{\uparrow \downarrow} (\hat{k}_1,\hat{k}_2,\hat{k}_3,\hat{k}_4) &= V^\ast_{\uparrow \downarrow} ({k}_1,{k}_2,{k}_3,{k}_4) 
 \, .
\end{align*}
 This symmetry is as well preserved in the fRG flow.
 The point-group symmetries give rise to the constraints
\begin{align*}
 C_\sigma (R_{\hat{O}}{k},R_{\hat{O}}{k}') &= C_\sigma ({k},{k}') \, , \\
 V_\uparrow (R_{\hat{O}}{k}_1,R_{\hat{O}}{k}_2,R_{\hat{O}}{k}_3,R_{\hat{O}}{k}_4) &= V_\uparrow ({k}_1,{k}_2,{k}_3,{k}_4) \, ,\\
 V_{\uparrow \downarrow} (R_{\hat{O}}{k}_1,R_{\hat{O}}{k}_2,R_{\hat{O}}{k}_3,R_{\hat{O}}{k}_4) &= V_{\uparrow \downarrow} ({k}_1,{k}_2,{k}_3,{k}_4) 
 \, ,
\end{align*}
 where $R_{\hat{O}} $ denotes the representation matrix corresponding to the element $\hat{O}$ in the point group $\mathcal{G}$.
 (For a more general discussion of point-group symmetries in \emph{multiband} models, see Ref.~\onlinecite{me-point-group}.)
  Here and throughout, we will assume that the parity operation \cite{disc-symm-Hubb} $\mathbf{k} \to - \mathbf{k} $ is contained in $\mathcal{G}$.

 Before we proceed further, let us briefly elaborate on the behavior under time reversal, which corresponds to the transformation
\begin{equation*}
 \psi_\sigma (x) \to i \sigma \bar{\psi}_{-\sigma} (-\tau,\mathbf{R}) \, , \quad
 \bar{\psi}_\sigma (x) \to i \sigma \psi_{-\sigma} (-\tau,\mathbf{R})
\end{equation*}
 (cf.\ Refs.~\onlinecite{eberlein_param,time-reversal,disc-symm-Hubb}). For the coupling functions, this translates to
\begin{align*}
 C_{-\sigma} ({k'}^{\cal T},k^{\cal T}) & \to C_\sigma ({k},{k}') \, , \\
 V_\uparrow (k^{\cal T}_4,k^{\cal T}_3,k^{\cal T}_2,k^{\cal T}_1) & \to V_\downarrow ({k}_1,{k}_2,{k}_3,{k}_4) \, , \\
 V_{\uparrow \downarrow} (k^{\cal T}_4,k^{\cal T}_3,k^{\cal T}_2,k^{\cal T}_1) & \to V_{\uparrow \downarrow} ({k}_1,{k}_2,{k}_3,{k}_4) \, ,
\end{align*}
  where $ k^{\cal T} = (k_0,-\mathbf{k}) $. 
 One can observe that, in the presence of OSP and parity invariance, the time-reversal operation acts on the interaction just as a spin flip.
 Clearly, a finite AF gap $\Delta (k)$ breaks time-reversal invariance in the quadratic part of the action and consequently
 also in the renormalized interaction.

 Note however that, in the absence of such a gap, SU(2) invariance would impose stronger constraints on the interaction than time-reversal symmetry,
 as the full SU(2) spin symmetry contains more than spin-flip invariance.
 As an approximation, one may hence enforce spin-flip invariance in the interaction without completely eliminating the signatures of the SU(2) breaking.
 This approximation will be further discussed in Appendix~\ref{sec:time-normal-app}.
\subsection{Time-normal approximation} \label{sec:time-normal-app}

  Typically, the bare interaction is time-reversal invariant. In the presence of OSP and parity inversion, this is equivalent to spin-flip invariance, i.e.,
\begin{align*} U_\uparrow^{\{s\}} (k_1,k_2,k_3) &= U_\downarrow^{\{s\}} (k_1,k_2,k_3) \, , \\ U_{\uparrow \downarrow}^{\{s\}} (k_1,k_2,k_3) &= U_{\uparrow \downarrow}^{\{\bar{s}\}} (k_2,k_1,k_1+k_2-k_3) 
\end{align*}
 holds, where $ \bar{s} = (s_2,s_1,s_4,s_3) $.
 In order to avoid confusion with a full SU(2) invariance in the interaction, we will henceforth speak of time-reversal invariance instead of spin-flip invariance. 
 This distinction is physically important, as time-reversal symmetry is a discrete one, while a SU(2) symmetry is continuous.

 At the mean-field level, time-reversal breaking interactions are absent.
 Such terms are, however, generated during the RG flow if the interaction is not of reduced-mean-field type (see Appendix~\ref{sec:improved-param}).
 In a first attempt to enter the AF phase within a fRG framework beyond mean field,
 neglecting the time-reversal breaking interactions may be a decent approximation.
In the following, we will call this the time-normal approximation.

 As laid out in Ref.~\onlinecite{mydissertation}, time-reversal breaking contributions to the interaction with zero momentum and frequency transfer
 can be shown to vanish in random-phase approximation in the case
 of the Hubbard model at half-filling. For the two-pocket model of Sec.~\ref{sec:chubu}, it appears unlikely that such terms should play a major role.
 Note that the time-normal approximation does not involve any approximations at the one-particle level, where the time-reversal symmetry is still broken.
 
 As will become clear in the following, the fRG flow in time-normal approximation
 still shows features that are not included in the mean-field picture.
 It seems an appealing strategy to first study these new features and to include time-reversal breaking contributions to the interaction in a further step.
 While the former is the subject of the remainder of this work, the latter will be left for future studies.

 In time-normal approximation, the remaining spin-symmetry group for the interaction is $ \mathrm{G}_\mathrm{t}= \mathrm{U}_z (1) \times \mathrm{Z}_2 $,
 which has a preferred axis, but no preferred orientation along this axis.
 The $ \mathrm{Z}_2 $ symmetry rules out such a preferred orientation. It stems from the spin-flip invariance enforced by omitting
 contributions to the renormalized interaction that would violate the conditions
\begin{align*}
 W_\uparrow^{\{s\}} (k_1,k_2,k_3,k_4) &= W_\downarrow^{\{s\}} (k_1,k_2,k_3,k_4) \, , \\
 W_{\uparrow \downarrow}^{\{s\}} (k_1,k_2,k_3,k_4) &= W_{\uparrow \downarrow}^{\{\bar{s}\}} (k_2,k_1,k_4,k_3) \, .
\end{align*}
 This $\mathrm{Z}_2$ invariance is, however, not enforced on the one-particle level.
[On the one-particle level, $ \mathrm{G}_\mathrm{t} $ would be equivalent to SU(2) in the sense that a $\mathrm{G}_\mathrm{t}$ symmetric one-particle
 Green's function is automatically SU(2) symmetric.]

  In addition to the time-normal approximation, triplet and anomalous pairing tendencies will be discarded here,
 since they appear to play a minor role in the presence of perfect nesting.
 In other words, the single-channel coupling functions $ \Phi_{\mathrm{SC},\uparrow \pm}$, $ \Phi_{\mathrm{SC} -} $,
 $ \Phi_{K -} $, $ \Phi_{\mathrm{axis}-} $, and $ \Phi_{xy-} $ are neglected. 
 The remaining interaction terms read as
\begin{align*} 
 W_\uparrow (k_1,k_2,k_3,k_4)^{\{s\}}  = & \, \tilde{\delta}_{\left\{ k_i \right\} } \left[ U_\uparrow^{\left\{ s \right\} }  (k_1,k_2,k_3) \right. \\
& - \frac{1}{2}\Phi_\mathrm{CDW}^{\left\{ s \right\} }  (k_1-k_3,k_1,k_2) \\ & 
- \frac{1}{2}\Phi_{z}^{\left\{ s \right\} }  (k_1-k_3,k_1,k_2) \\ &
 + \frac{1}{2}\Phi_\mathrm{CDW}^{\left\{ \tilde{s} \right\} }  (k_3-k_2,k_1,k_2) \\ &
 + \left. \frac{1}{2}\Phi_{z}^{\left\{ \tilde{s} \right\} }  (k_3-k_2,k_1,k_2)  \right] \, ,
\end{align*}
\begin{align*}
 W_{\uparrow \downarrow}^{\{s\}} (k_1,k_2,k_3,k_4)  =  & \, \tilde{\delta}_{\left\{ k_i \right\} } \left[ U_{\uparrow \downarrow}^{\left\{ s \right\} }  (k_1,k_2,k_3) \right. \\ &+ \Phi_\mathrm{sing}^{\left\{ s \right\} }  (k_1+k_2,k_1,k_3) \\
&+ \Phi_{xy +}^{\left\{ s \right\} }  (k_3-k_2,k_1,k_2) \\ &
- \frac{1}{2} \Phi_\mathrm{CDW}^{\left\{ s \right\} }  (k_1-k_3,k_1,k_2) \\
&+ \left. \frac{1}{2} \Phi_z^{\left\{ s \right\} }  (k_1-k_3,k_1,k_2) \right]
\end{align*}
and
\begin{equation*}
 W_\downarrow^{\{s\}} (k_1,k_2,k_3,k_4)  = W_\uparrow^{\{s\}} (k_1,k_2,k_3,k_4)  \, .
\end{equation*}
 Consequently, one now has $ W_{\pm \uparrow} = W_{\pm \downarrow} \equiv W_{\pm} $ for the shorthand notations introduced in Eq.~(\ref{eqn:pm-shorthands}).
 It can easily be verified, that the time-normal approximation gives rise to
\begin{equation*}
  W_{\pm}^{\{s\}} (k_1,k_2,k_3,k_4)  =W_{\pm}^{\{\bar{s}\}} (k_2,k_1,k_4,k_3) \, . 
\end{equation*}
 The flow equation for the single-channel coupling functions in time-normal approximation
 can be cast into a simple form, where the loops enter in the spin-symmetrized combinations
\begin{align*}
 I_\mathrm{eq}^{\{s'\}} (l,p) & = \frac{1}{2} \left[
 G_\uparrow^{s'_1 s'_2} (p-l/2) \, G_\uparrow^{s'_3 s'_4} (p+l/2) \right. \\ & \qquad  \left.
+ G_\downarrow^{s'_1 s'_2} (p-l/2) \, G_\downarrow^{s'_3 s'_4} (p+l/2) \right]\, ,
\end{align*}
\begin{align*}
 I_\mathrm{op}^{\{s'\}} (l,p) & = \frac{1}{2} \left[
 G_\uparrow^{s'_1 s'_2} (p-l/2) \, G_\downarrow^{s'_3 s'_4} (p+l/2)  \right. \\ & \qquad  \left.
+ G_\downarrow^{s'_1 s'_2} (p-l/2) \, G_\uparrow^{s'_3 s'_4} (p+l/2) \right]\, ,
\end{align*}
\begin{align*}
 J_\mathrm{op}^{\{s'\}} (l,p) & = \frac{1}{2} \left[
 G_\uparrow^{s'_1 s'_2} (l/2+p) \, G_\downarrow^{s'_3 s'_4} (l/2-p)  \right. \\ & \qquad  \left.
 + G_\downarrow^{s'_1 s'_2} (l/2+p) \, G_\uparrow^{s'_3 s'_4} (l/2-p) \right]\, .
\end{align*}
 
 For the singlet-pairing channel, one obtains the flow equation
\begin{align*}
 \dot{\Phi}_\mathrm{sing}^{ \{s\}}  (l,q,q') &= - \frac{1}{2} \sum_{\{s'\}} \int \! d'p \,
  \dot{J}_\mathrm{op}^{\{ s'\}} (l,p) \\ & \quad \times W_{\uparrow \downarrow}^{s_1,s_2,s'_1,s'_3} (q,l-q,l/2+p,l/2-p) \\
& \quad \times \left[ W_{\uparrow \downarrow}^{s'_2,s'_4,s_3,s_4} (l/2+p,l/2-p,q',l-q') \right. \\
  & \quad \left. + W_{\uparrow \downarrow}^{s'_4,s'_2,s_3,s_4} (l/2-p,l/2+p,q',l-q') \right]
\end{align*}
 and for the in-plane magnetic channel
\begin{align*}
  \dot{\Phi}_{xy +}^{\left\{ s \right\}} (l,q,q')  &= -  \sum_{\left\{ s' \right\}} \int \!\! d' p \,
\dot{I}_\mathrm{op}^{\{ s'\}} (l,p) \\ 
 & \quad \times W_{\uparrow \downarrow}^{s'_4,s_2,s_3,s'_1} (p+l/2,q',l+q',p-l/2)\\
 & \quad \times W_{\uparrow \downarrow}^{s_1,s'_2,s'_3,s_4} (q,p-l/2,p+l/2,q-l) \, .
\end{align*}
  For the CDW channel, the flow equation reads as
\begin{align*} 
  \dot{\Phi}_\mathrm{CDW}^{\left\{ s \right\}} (l,q,q') &= -  \sum_{\left\{ s' \right\}} \int \!\! d' p \,
\dot{I}_\mathrm{eq}^{\{ s'\}} (l,p) \\
& \quad \times W_-^{s'_2,s_1,s'_3,s_3} (p-l/2,q,p+l/2,q-l)\\
&  \quad \times W_-^{s'_4,s_2,s'_1,s_4} (p+l/2,q',p-l/2,q'+l) \, ,
\end{align*}
  and for the $ S^2_z $ channel one gets
\begin{align*}
  \dot{\Phi}_{z}^{\left\{ s \right\}} (l,q,q') &= -  \sum_{\left\{ s' \right\}} \int \!\! d' p \,
\dot{I}_\mathrm{eq}^{\{ s'\}} (l,p) \\
 & \quad \times W_+^{s'_2,s_1,s'_3,s_3} (p-l/2,q,p+l/2,q-l) \\
 & \quad \times W_+^{s'_4,s_2,s'_1,s_4} (p+l/2,q',p-l/2,q'+l) \, .
\end{align*}
 In the following, the self-energy will be decomposed into its spin-symmetric and spin-antisymmetric parts
\begin{equation*} 
 \Sigma_\pm^{s_1 s_2} (k) = \frac{1}{2} \left[\Sigma_\uparrow^{s_1 s_2} (k) \pm \Sigma_\downarrow^{s_1 s_2} (k) \right] \, ,
\end{equation*} 
 which flow according to
\begin{align*}
 \dot{\Sigma}_\pm^{s_1 s_2} (k) & = - \frac{1}{2} \sum_{s'_1 s'_2 } \int \!\! d' p \, \left[ S_\uparrow^{s'_1 s'_2 } (p) \pm S_\downarrow^{s'_1 s'_2 } (p) \right] 
 \\
 & \qquad \times W_\mp^{s_1 s'_2 s_2 s'_1} (k,p,k,p) \, .
\end{align*} 
 Expressed in terms of the quantities defined in Appendix~\ref{sec:AFM-symm},
 $\Sigma_+$ contains $ \operatorname{Re} \epsilon_\mathrm{s} $, $ \operatorname{Im} \epsilon_\mathrm{a} $ and $ \operatorname{Im} \Delta $ and $ \Sigma_- $ contains $\operatorname{Re} \Delta $,
 $ \operatorname{Im} \epsilon_\mathrm{s} $ and renormalizations of $ \operatorname{Re} \epsilon_\mathrm{a} $.

 \subsection{Exchange parametrization} \label{sec:AFM-ex-para}
 As in Refs.~\onlinecite{husemann_09,HGS-freq-dep,giering-new}, one may now resort to an exchange parametrization.
 The formalism presented here has been adapted from Ref.~\onlinecite{giering-new}, where the Hubbard model has been studied in the symmetric phase.

 Two slowly varying form factors are  already encoded in the Nambu indices.
 They can be attributed to the two irreducible representations of the $ \mathrm{Z}_2$ group, which correspond to basis vectors that are even or odd under a
 Nambu-index flip.
 We will refer to them as trivial and sign-changing form factors, respectively.
 For the two-pocket model of Sec.~\ref{sec:chubu}, they correspond 
to $s$-wave or $ s^\pm$-wave, respectively. For the Hubbard model, the trivial form factor is of  $s$-wave and the sign-changing one of $d_{x^2-y^2} $-wave type.%
\footnote{In the latter case of the Hubbard model, the magnetic Brillouin zone should be centered around $ (0,\pi) $ for simplicity.}
 Although one may in principle include more form factors, these two slowly varying ones appear suitable for low-energy considerations.

 In the present case, there is not only one unique way to perform an exchange parametrization.
 In particular, the dependence of the interaction on the Nambu indices can be treated in various ways and the dependence on the weak momenta and frequencies 
 can either be taken into account within a form-factor expansion as in Ref.~\onlinecite{husemann_09} and subsequent works, or it may be projected 
 to a single point. In the following, the latter strategy will be pursued.
 Moreover, it appears sensible to first resort to an exchange parametrization which does not contain approximations on the Nambu-index dependence.
 Such approximations can then still be devised at a later stage in agreement with the symmetries of the system.
 On a formal level, a single-channel coupling function $ \Phi_P$ is approximated
 by the product of fermion-boson vertices $ g_\alpha (q,l) $ and exchange propagators 
 $ P^{\{s\}}_{\alpha \beta} (l) $, i.e.\
\begin{equation*}
 \Phi_P^{\{s\}} (l,q,q') \approx \sum_{\alpha \beta} g_\alpha (q,l) \,g_\beta (q',l) \, P^{\{s\}}_{\alpha \beta} (l) \, .
\end{equation*}
 The indices $\alpha$ and $\beta$  correspond to bosonic flavors here.
 In the following, only fermion-boson vertices with a trivial momentum and frequency dependence will be taken into account.
 Since $\alpha$ and $\beta$ then take on only one value, they will be suppressed from the notation in the following.
 Normalizing the momentum- and frequency-independent fermion-boson vertices to unity then gives
\begin{equation*}
 \Phi_P^{\{s\}} (l,q,q') \approx P^{\{s\}} (l) \, ,
\end{equation*}
 i.e.\ for each combination of Nambu indices, the coupling function of a particular channel can then
 be attributed to a bosonic propagator.
 The above-mentioned trivial and sign-changing form factors then come into play if the exchange propagators are parametrized further,
 which will be addressed in the following.

 More precisely, one may choose
\begin{equation*}
 \Phi_\mathrm{sing}^{\{s\}} (l,q,q') \approx D^{\{s\}} (l) = \mathcal{P}_\mathrm{pp} \left[\Phi_\mathrm{sing}^{\{s\}}  \right] (l)
\end{equation*}
 with the projection rule
\begin{equation*}
\mathcal{P}_\mathrm{pp} \left[\Phi \right] (l) = \Phi (l,l/2,l/2)
\end{equation*}
 for particle-particle channels.
 Note that the weak momentum and frequency arguments of $\Phi$ are chosen in such a way that unique symmetry constraints on the exchange propagators
 result from the respective constraints on $\Phi$. (For further details, see Ref.~\onlinecite{mydissertation}.)
 For the particle-hole channels, the bosonic propagators are defined likewise according to
\begin{align*}
 \Phi_\mathrm{CDW}^{\{s\}} (l,q,q') & \approx N^{\{s\}} (l) = \mathcal{P}_\mathrm{ph} \left[ \Phi_\mathrm{CDW}^{\{s\}}  \right] (l)\, ,\\
 \Phi_{xy +}^{\{s\}} (l,q,q') & \approx M_{xy}^{\{s\}} (l) = \mathcal{P}_\mathrm{ph} \left[ \Phi_{xy+}^{\{s\}}  \right] (l) \, ,\\
 \Phi_z^{\{s\}} (l,q,q') & \approx M_z^{\{s\}} (l) = \mathcal{P}_\mathrm{ph} \left[ \Phi_z^{\{s\}}  \right] (l)\, .
\end{align*}
 The projection rule
\begin{equation*}
\mathcal{P}_\mathrm{ph} \left[\Phi \right] (l) = \Phi (l,l/2,-l/2)
\end{equation*}
 for the particle-hole channels differs from $\mathcal{P}_\mathrm{pp}$  by a minus sign in the last argument of $\Phi$,
 which ensures compatibility with the symmetries.
 
 Note that in contrast to Refs.~\onlinecite{husemann_09,HGS-freq-dep,giering-new}, not only the weak frequency dependencies, but also the weak momentum dependencies
 are projected to a single point through the above projection rule. For this work, this seems to be an adequate choice, since the two-pocket model
  of Sec.~\ref{sec:chubu} can be regarded in the light of a gradient expansion around the
 centers of hole and electron pockets. The physics of the Hubbard model at van Hove filling is dominated by the vicinity of
 the saddle points $ (0,\pi)$ and $ (\pi,0) $ of its dispersion and therefore the projection rule presented here may also be applicable in that case.
 For $l=0$, the weak frequencies and momenta are only considered in leading (zeroth) order in a gradient expansion around these hot spots.

 Of course, different projection rules can be applied for lattice models, such as the form-factor expansion rule of 
Refs.~\onlinecite{husemann_09,HGS-freq-dep,giering-new} or the Fermi surface projection of Refs.~\onlinecite{eberlein-sssu,eberlein-phd,eberlein-repu}. 
These projection rules 
 all comply with the OSP, the PHS, and the Pauli principle.

 Let us now assume that the bare interaction is time-reversal invariant and only depends on the Nambu indices and not on the momenta.
 Both the bare interactions of the Hubbard model and of the two-pocket model considered in this work are of this type.
 In terms of the above defined exchange propagators, the multichannel coupling functions $ W_{\uparrow \downarrow} $,
 $ W_+ $, and $ W_-$ in time-normal approximation read as
\begin{align*}
%W_\uparrow^{\{ s\}} (k_1,k_2,k_3)  & = U_\uparrow^{\{s\}} 
%+ \frac{1}{2} \left[ N^{\{ \tilde{s}\}}(k_3-k_2) + M_z^{\{ \tilde{s}\}}(k_3-k_2) \right] \\ & \quad
%- \frac{1}{2} \left[ N^{\{ s\}}(k_1-k_3) + M_z^{\{ s\}}(k_1-k_3) \right] 
%\, , \\
 W_{\uparrow \downarrow}^{\{ s\}} (k_1,k_2,k_3) &= 
U_{\uparrow \downarrow}^{\{s\}} 
+ D^{\{ s\}}(k_1+k_2) + M_{xy}^{\{ s\}} (k_3-k_2) \\ & \quad
 + \frac{1}{2} \left[ M_z^{\{ s\}}(k_1-k_3) - N^{\{ s\}}(k_1-k_3) \right] \, ,
\end{align*}
\begin{align*}
W_+^{\{ s\}} (k_1,k_2,k_3)  & = U_{\uparrow \downarrow}^{\{s\}} - U_\uparrow^{\{s\}} 
+ D^{\{ s\}}(k_1+k_2)  \\ & \quad + M_{xy}^{\{ s\}} (k_3-k_2)
+ M_z^{\{ s\}}(k_1-k_3) \\ & \quad 
- \frac{1}{2} \left[ N^{\{ \tilde{s}\}}(k_3-k_2) + M_z^{\{ \tilde{s}\}}(k_3-k_2) \right] \, ,
\end{align*}
\begin{align*}
W_-^{\{ s\}} (k_1,k_2,k_3)  & = U_{\uparrow \downarrow}^{\{s\}} + U_\uparrow^{\{s\}} 
+ D^{\{ s\}}(k_1+k_2) \\ & \quad + M_{xy}^{\{ s\}} (k_3-k_2)
- N^{\{ s\}}(k_1-k_3) \\ & \quad 
+ \frac{1}{2} \left[ N^{\{ \tilde{s}\}}(k_3-k_2) + M_z^{\{ \tilde{s}\}}(k_3-k_2) \right] \, ,
 \end{align*}
 where we have assumed that the bare interaction is featureless in momentum space.
\subsection{Nambu-normal approximation}
 
 At this level, a direct implementation of the fRG flow equations with Nambu-index-dependent exchange propagators would be quite costly.
 Therefore, we will resort to additional approximations.
 First, let us point out that the Nambu-index dependence of the interaction may as well be attributed to the fermion-boson vertices instead of the
 exchange propagators. On a formal level, this corresponds to a product ansatz for the $ P^{\{s\}} $, i.e.,
\begin{align*}
D^{\{ s\}}(l) &= \sum_{m,m'=0,1} \, \sum_{n,n'=\pm} g_{s_1 s_2}^{m,n}\, g_{s_3 s_4}^{m',n'} \, D_{m m'}^{n n'} (l) \, , \\
  M_{xy}^{\{ s\}}(l) &= \sum_{m,m'=0,1} \, \sum_{n,n'=\pm} g_{s_1 s_4}^{m,n} \, g_{s_3 s_2}^{m',n'} \, \left(M_{xy} \right)_{m m'}^{n n'} (l) \, , \\
  M_z^{\{ s\}}(l) &= \sum_{m,m'=0,1} \, \sum_{n,n'=\pm} g_{s_1 s_3}^{m,n} \, g_{s_2 s_4}^{m',n'} \, {M_z}_{m m'}^{n n'} (l) \, ,\\
  N^{\{ s\}}(l) &= \sum_{m,m'=0,1} \, \sum_{n,n'=\pm} g_{s_1 s_3}^{m,n} \, g_{s_2 s_4}^{m',n'} \, N_{m m'}^{n n'} (l) \, .
\end{align*}
  The Nambu-index-dependent part of the fermion-boson vertices can then be factorized as $ g_{s s'}^{n,m} = \tau_{s s'}^m \, f_{s }^n $, 
 where $\tau_{s s'}^m$ accounts for the
 ordering momentum and where $ f^s_n $ corresponds to a form factor.
 More precisely, the trivial and sign-changing form factors read as $ f^{s }_+ = 1 $ and $  f^{s }_- = s $, respectively.
 Again note that, strictly speaking, the assignment of $\tau^0$ to ordering momenta around $0$ and of $\tau^1$ to ordering momenta around $(\pi,\pi)$
 only holds in the sense of a gradient expansion.
 In the following, let us assume the bare interaction to be momentum conserving and featureless in momentum space, i.e., to correspond to
 a Hubbard on-site term of strength $U$.
 
 We now neglect interactions terms that conserve momentum only up to $\mathbf{Q}$. On the one-particle level, momentum conservation then will of course still be violated.
 In addition, we restrict ourselves to trivial form factors and neglect contributions with $\tau_{s s'}^0 $ in the particle-hole channel and with
 $\tau_{s s'}^1 $ in the particle-particle channel, which are presumably of minor importance. 

 For a more detailed discussion of these approximations, which we call \emph{Nambu normal}, we refer to Ref.~\onlinecite{mydissertation}.
 They finally yield the flow equations~(\ref{eqn:red-D})--(\ref{eqn:red-se}) and the Ward identity~(\ref{eqn:red-WI}).
\end{appendix}
\bibliography{biblio}
\end{document}